\PassOptionsToPackage{usenames,dvipsnames}{xcolor}
\documentclass[USenglish,oneside,twocolumn]{article}

\usepackage[utf8]{inputenc}
\usepackage[big]{dgruyter_NEW}

\usepackage{graphicx}
\usepackage{algorithm}
\usepackage{amsmath,amssymb,amsfonts}
\usepackage{amsthm}
\usepackage{textcomp}
\usepackage{xcolor}
\usepackage{paralist}
\usepackage{tabularx}
\usepackage{textcomp,booktabs}
\usepackage{numprint}
\npthousandsep{,}
\npdecimalsign{.}
\usepackage{advdate}

\def\BibTeX{{\rm B\kern-.05em{\sc i\kern-.025em b}\kern-.08em
    T\kern-.1667em\lower.7ex\hbox{E}\kern-.125emX}}

\makeatletter
\def\plist@algorithm{Alg.\space}
\makeatother

\newtheorem{claim}{Claim}

\theoremstyle{definition}
\newtheorem{definition}{Definition}[section]

\newtheoremstyle{theoremdd}
  {0pt}
  {0pt}
  {}
  {0pt}
  {\bfseries}
  {. }
  { }
  {\thmname{#1}\thmnumber{ #2}\thmnote{ (#3)}}

\theoremstyle{theoremdd}
\newtheorem{myrestriction}{Restriction}[section]

\newcommand{\usd}[1]{\numprint {#1}}

\providecommand{\keywords}[1]{{\bf Primary Keyword}: #1}
\newcommand{\ltcToLito}[1]{\FPeval{\result}{round(#1*100000000,0)}}

\newcommand{\newcontent}[1]{\textcolor{black}{{#1}}}

\newcommand{\ltcPrice}{239.40} 
\newcommand{\ltcToUSD}[1]{\FPeval{\result}{round(#1*\ltcPrice,4)}}
\newcommand{\roundtwo}[1]{\FPeval{\result}{round(#1,2)}}
\newcommand{\roundfour}[1]{\FPeval{\result}{round(#1,6)}}

\usepackage{datatool}
\DTLsetseparator{,}

\DTLloaddb{bbcdata}{bbc_stats.txt}

\DTLforeach*
{bbcdata}
{\entries=entries,\size=size, \totalarticles=total_articles, \cost=total_cost, \avgcost=avg_cost, \avgswift=avg_swift_thr, \avgonchain=avg_onchain_thr}

\cclogo{\includegraphics{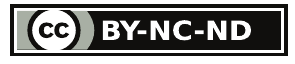}}


\journalname{Proceedings on Privacy Enhancing Technologies}
\DOI{Editor to enter DOI}
  \startpage{1}
  \received{..}
  \revised{..}
  \accepted{..}

  \journalyear{..}
  \journalvolume{..}
  \journalissue{..}

\begin{document}
\begin{filecontents*}{bbc_stats.txt}
total_articles, size, entries, avg_swift_thr, avg_onchain_thr, avg_cost, total_cost
53823, 51208863, 134, 3469.19661936, 1245.45321813, 0.000011, 0.5751316
\end{filecontents*}

\author*[1]{Ruben Recabarren}
\author[2]{Bogdan Carbunar}
\affil[1]{Florida International University, Miami, FL 33199, E-mail: recabarren@gmail.com}
\affil[2]{Florida International University, Miami, FL 33199, E-mail: carbunar@gmail.com}

\title{\huge Toward Uncensorable, Anonymous and Private Access Over Satoshi Blockchains}
\runningtitle{Toward Uncensorable, Anonymous and Private Access Over Satoshi Blockchains}

\begin{abstract}
{Providing unrestricted access to sensitive content such as news and software is difficult in the presence of adaptive and resourceful surveillance and censoring adversaries. In this paper we leverage the distributed and resilient nature of commercial Satoshi blockchains to develop the first provably secure, censorship resistant, cost-efficient storage system with anonymous and private access, built on top of commercial cryptocurrency transactions. We introduce max-rate transactions, a practical construct to persist data of arbitrary size entirely in a Satoshi blockchain. We leverage max-rate transactions to develop UWeb, a blockchain-based storage system that charges publishers to self-sustain its decentralized infrastructure. UWeb organizes blockchain-stored content for easy retrieval, and enables clients to store and access content with provable anonymity, privacy and censorship resistance properties.\\
We present results from UWeb experiments with writing \FPeval{\result}{round((217+\size/1000000),2)}\result~MB of data into the live Litecoin blockchain, including 4.5 months of live-feed BBC articles, and 41 censorship resistant tools. The max-rate writing throughput (183 KB/s) and blockchain utilization (88\%) exceed those of state-of-the-art solutions by 2-3 orders of magnitude and broke Litecoin's record of the daily average block size. Our simulations with up to 3,000 concurrent UWeb writers confirm that UWeb does not impact the confirmation delays of financial transactions.}
\end{abstract}
\maketitle

\section{Introduction}

Basic human rights continue to be eroded around the world through fine-grained monitoring of Internet access~\cite{NSA, govMonSN}, and restricted access to information~\cite{BlockedInChina, BlockedInRussia} that includes select news, software artifacts~\cite{ChinaBan, USban} and even research~\cite{GuardianCensorship, ChinaResearchCensor}. Traditional solutions such as private access services, e.g., VPNs, centralize user access and were shown to keep access records~\cite{vpnLog} and share them with censors~\cite{vpnFlaws,vpnChina,turboVPN}. Tor and its onion routing and hidden services have well known limitations that simplify deanonymization not only for governments but also for some corporations~\cite{TorDeanon}.

In this paper we leverage the distributed nature and substantial collateral damage inflicted by blocking cryptocurrency blockchains, to develop a censorship-resistant storage system that provides private and secure data access, by embedding data into cryptocurrency transactions.

Blockchains have been advertised as a platform for distributed services~\cite{Blockchain.IBM, Blockchain.amazon, Blockchain.Microsoft, MaidSafe, FileCoin, Sia, LTI, KKASGJGF18, DXRSW17, ANSF16, ASNF17, TD17, LSTKKN17, LYCXZ17, SBHD17, CC17, HL16, SBRS16, wachs2014censorship, MJSPK14}, including for censorship resistance~\cite{K19, Z18, RC19, MMK20}. For instance, early efforts have used blockchains to post sensitive articles, e.g., on vaccine-related offenses of a biotechnology company, and bypass government censorship~\cite{K19, Z18}. The ideal blockchain-storage solution should however (1) support more frequent writing needs and larger content, and (2) rely on a blockchain that is completely distributed, and sufficiently popular to inflict unpalatable collateral damage to censors and to have a mining hashrate high enough to make it difficult to launch majority attacks.

In this paper we develop techniques to significantly extend the amount of data that can be stored on Satoshi blockchains, i.e., Bitcoin and its variants. At the time of writing, Satoshi cryptocurrencies have the highest market capitalization thus can inflict the highest collateral damage to would-be censors: the total market cap of the top 50 Satoshi cryptocurrencies exceeds \$1 trillion~\cite{coinmarketcap}.

Our quest introduces however a new set of requirements that include (1) practicality in terms of storage throughput, goodput and cost, (2) efficient retrieval and access of data stored among millions of transactions, e.g., 615 million transactions in Bitcoin~\cite{totaltxnsbitcoin}, 60 million in Litecoin~\cite{totaltxnslitecoin}, and (3) satisfaction of constrains imposed by Satoshi bockchains, e.g., on the transaction size, the number of input and output counts, the transaction fees, the number of unconfirmed transactions and confirmation times. The ideal solution should also provide fully on-chain storage, to avoid the censorship and surveillance vulnerabilities introduced by hybrid storage solutions~\cite{ANSF16, ASNF17}.

In addition, non-segwit Satoshi transactions are vulnerable to {\it integrity attacks} (Appendix~\ref{appendix:attacks}), also known as malleability attacks \cite{bitcoinMalleability}. Of particular concern are integrity attacks where transactions are modified within data portions unprotected by cryptographic signatures.

To simultaneously address the blockchain-writing constraints, in this paper we introduce {\it max-rate transactions}, data-storage constructs that use cryptocurrency transactions to prevent integrity attacks, and further optimize the writable space within the building blocks of a single transaction and the chaining of data-embedding transactions to maximize the amount of data written per time unit, thus minimize data access latency.

To enable efficient search and access of blockchain-stored data we build on max-rate transactions to introduce UWeb, the first practical, entirely on-chain storage system that efficiently organizes blockchain-stored data in directory-inspired structures. UWeb's main use case is the one-to-many distribution of popular but sensitive content, e.g., news and software.

We prove that an all-powerful monitor cannot distinguish UWeb users from regular cryptocurrency users and cannot determine what data they access. Further, blocking access to UWeb would deprive the economy of the censored region from access to a financial market whose capitalization exceeds \$1 trillion.

We use max-rate transactions to write a total of \FPeval{\result}{round((217+\size/1000000),2)}\result ~MB of content considered sensitive by many censors, i.e., BBC articles and censorship evading software, into the Litecoin blockchain. Our experiments reveal the practicality of the proposed solutions, that achieved an aggregate throughput of 183 KB/s. They also increased the average Litecoin daily block size to 206KB, breaking Litecoin's lifetime record.

Our writing experiments had no effects on the confirmation times of regular financial transactions. We further confirmed this through simulations with up to 3,000 concurrent UWeb writers (1.14GB issued in a 4 hour interval) and up to 10 times more financial transactions.

All the data written in our experiments is available for free public access in the Litecoin blockchain by inspecting the spending of the Litecoin address LZAhHQjxf6dQaTxTAK7g1wTz3hZRXX5MkG.

In summary, our contributions are the following:

\begin{compactitem}

\item
{\bf Max-rate transactions}.
We develop the first blockchain-writing constructs that, for the lowest price, and with a single input address, provably achieve a storage throughput asymptotic to the available bandwidth, and a goodput that approaches the theoretic limit. We prove that max-rate transactions are standard and prevent integrity attacks.

\item
{\bf UWeb}.
We introduce the first practical, entirely on-chain, secure and private storage system that leverages max-rate transactions to efficiently discover, recover and reconstruct content of interest embedded among hundreds of millions of transactions. We prove that UWeb provides users with anonymous and private access, and censorship resistance.

\item
{\bf Litecoin mainnet experiments}.
We show through experiments with writing more than 268MB of data in the live Litecoin mainnet (not testnet) that UWeb achieved 2-3 orders of magnitude improvements in throughput and block utilization, and wrote 4-5 orders of magnitude more data from a single funding address, when compared to state-of-the-art solutions.

\end{compactitem}
\vspace{-15pt}
\section{Background and Related Work}
\label{sec:related}

\noindent
{\bf Cryptocurrency Transactions}.
We model a cryptocurrency transaction as a tuple $\tau = (v, f, I_T, O_T, w, l_o)$, where $v$ is the version number, $f$ is a flag that indicates the presence of witness data, $I_\tau$ is a list of inputs, $O_\tau$ is a list of outputs, $w$ an optional block of witness data, and $l_o$ is the transaction lock time. $|\tau|$, used to denote the size of $\tau$, is the total size of its components. The \textit{transaction id} of $\tau$ is the double SHA256 hash of $\tau$'s concatenated components.

The list of inputs is $I_\tau = \{I_x | 0 \leq x \leq c\}$, where $c$ is the total count of inputs used to prefix the list. An input is a tuple $I_x = (p_h, p_i, s_i, z)$. $p_h$ is the id of the previous transaction that contains the funding output for $I_\tau$, $p_i$ is the output's index in the transaction with id $p_h$, $s_i$ is a \textit{script} (called the \textit{scriptSig} script) used to verify that a user is authorized to spend the balance from transaction $p_h$ and index $p_i$, and $z$ is a sequence number related to the transaction lock time.

Similarly, the count-prefixed output of transaction $\tau$ is defined as $O_\tau = \{O_x : 0 \leq x \leq c\}$, where $c$ is the number of outputs used as prefix for list, and an output $O_x$ is a tuple $O_x = (s_o, o_v)$. $s_o$ is the \textit{scriptPubKey} script, and $o_v$ is the output value to be transferred from the sum of values specified in the transaction $\tau$ list of inputs. A transaction $\tau$ is invalid if the sum of the values from the inputs is smaller than or equal to the sum of the outputs. The balance after subtracting the output values is considered to be the {\it miner fee}. We note that a zero-fee transaction will not be mined by modern pools~\cite{minerfees} and it may not be broadcast by default configured Satoshi-compliant nodes~\cite{minrelayfee}.

A transaction that has not yet been mined into a block is said to be {\it unconfirmed}. Further, a pair of transactions where one transaction spends the other's value are said to be {\it chained}. Using this basic knowledge of Satoshi networks, previous efforts have designed sub-optimal or insecure data insertion solutions that we summarize next.

\noindent
{\bf Apertus}.
Kaminsky~\cite{kaminskyPayToHash} proposed the first blockchain-writing solutions, that use the output address bytes in the {\it scriptPubKey} to store data: the Pay-to-PubkeyHash (p2pkh) and Pay-to-Script-Hash (p2sh) techniques. Apertus~\cite{Apertus} uses p2pkh writing that overwrites the 20 bytes of a destination address to store arbitrary data.

\noindent
{\bf Catena and Blockstack}.
Catena~\cite{TD17} and Blockstack~\cite{ANSF16} introduce inexpensive solutions for small payloads. These systems use an OP\_RETURN based writing, to mark a transaction as invalid, and output un-spendable transactions that are immediately prunable from the un-spent transaction set. This contract allows for writing 80 bytes after the OP\_RET opcode. Since standard transactions can only carry one OP\_RET output \cite{opretcheck}, it is impossible to improve the efficiency of this construct, thus limiting Catena and Blockstack to sub-optimal blockchain utilization.

\noindent
{\bf Blockchain-Based Censorship Resistance}.
Early attempts have written a few sensitive articles (e.g., on vaccination misbehaviors of a biotech company) on the Ethereum blockchain to avoid censorship~\cite{K19, Z18}. More rigorous, academic efforts include Tithonus~\cite{RC19} and MoneyMorph~\cite{MMK20}. Tithonus~\cite{tithonus} provides solutions that allow censored clients to surreptitiously embed paid requests for sensitive content into Bitcoin transactions and also for uncensored services to redeem the payment when they embed the requested content into transactions. MoneyMorph~\cite{MMK20} further designs rendezvous protocols over the blockchains of commercial cryptocurrencies to bootstrap censorship resistant communications. When compared against Bitcoin, Monero and Ethereum, Zcash provides MoneyMorph with the best bandwidth per transaction and the lowest cost.

Table~\ref{tables:solutions:comparison} in $\S$~\ref{sec:evaluation:state} provides details of the comparison of the max-rate transactions that we develop in this paper against state-of-the-art blockchain-writing solutions (Apertus, Catena, MoneyMorph and Tithonus). In addition, our solutions embed 4.6MB in the blockchain in a single mining event, improving significantly over MoneyMorph that embedded 20 bytes into a Bitcoin transaction, 20 bytes in Ethereum, 640 in Monero and 1,148 in a Zcash transaction, and also Tithonus that was able to embed 1,635 bytes into a Bitcoin transaction. Max-rate transactions thus achieve a storage throughput that improves on existing solutions by 3-4 orders of magnitude, and a blockchain utilization that improves on existing solutions by 2-4 orders of magnitude. In addition, max-rate transactions address the Tithonus vulnerability to integrity attacks.

Further, UWeb improves on existing blockchain-writing solutions by providing novel, on-chain-only techniques to organize and update content stored in the blockchain for efficient retrieval, and to access content with provable privacy and censorship resilience.

\noindent
{\bf Staged Transactions}.
Unlike previous solutions that overload portions of the output scripts, {\it staged transactions}, documented by Todd~\cite{inputWritingPeterTodd}, use the inputs section, the largest field in typical transactions. Unlike output writing where only one transaction is needed, input overloading requires the use of a pair of transactions: A {\it funding} transaction with a p2sh output that specifies the hash of its redeeming script, and a {\it spending} transaction, whose input script provides a {\it redeemScript} that satisfies the funding transaction's conditions, and stores the actual arbitrary data. Tithonus~\cite{tithonus}, a censorship resilient system, introduces variations of this technique, that are however vulnerable to output and input script modification attacks, see $\S$~\ref{sec:model:adversary}. In $\S$~\ref{sec:evaluation:state} we show that our proposed blockchain-writing techniques are significantly more efficient than state-of-the-art solutions.

\noindent
{\bf Blockchain Based Services}.
Blockchains have been used to store arbitrary user data~\cite{SBRS16, MJSPK14, Sia, MaidSafe, FileCoin}, 
sensitive data~\cite{KKASGJGF18}, including medical records~\cite{DXRSW17}, to provide a decentralized PKI system~\cite{ANSF16,ASNF17}, data provenance for clouds~\cite{LSTKKN17}, a privacy preserving domain name system~\cite{wachs2014censorship}, data integrity assurances for cloud-based IoT~\cite{LYCXZ17,SBHD17}, to implement content sharing and access control for social networks~\cite{CC17}, and to secure the BGP and DNS infrastructure~\cite{HL16}.

In this paper we develop a censorship-resistant blockchain-based storage service that uses decentralized blockchains to avoid single points of censorship~\cite{Blockchain.IBM, Blockchain.amazon, Blockchain.Microsoft}. Further, to avoid majority attacks~\cite{Majority.attack}, e.g., through opportunistic renting of mining equipment using sites like nicehash~\cite{nicehash}, we build UWeb on Satoshi blockchains, thus we improve on solutions based on custom-made blockchains~\cite{MJSPK14} or blockchains that have small gossip networks and mining infrastructure~\cite{Sia, MaidSafe, FileCoin}.

Solutions like Blockstack~\cite{ANSF16,ASNF17} implement a hybrid blockchain/cloud approach, where the data is stored in traditional cloud storage, e.g. Amazon S3, and only a cryptographic pointer to this data is stored in the blockchain. UWeb instead stores all data on-chain, including metadata, e.g., directory structure, public key certificates. This endows UWeb with resistance to censorship and privacy compromise, since an adversary can no longer obtain cooperation from the cloud provider or correlate blockchain and cloud accesses.

\vspace{-15pt}

\section{Model and Problem Definition}
\label{sec:model}

\vspace{-5pt}

We first describe the UWeb ecosystem then define the adversary model and the problems that we seek to solve.

\vspace{-5pt}

\subsection{System Model}
\label{sec:model:system}

\begin{figure}
\centering
\includegraphics[width=0.49\textwidth]{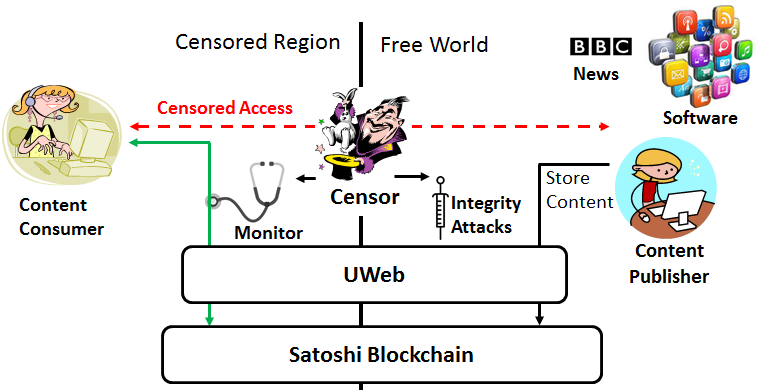}
\caption{System and adversary model. A client in the censored region cannot directly access sensitive services and news. Instead, content publishers embed content on commercial blockchains; clients access data with privacy and anonymity through standard Satoshi clients. The adversary can censor select client communications, monitor communications and perform integrity attacks.}
\label{fig:system:model}
\vspace{-15pt}
\end{figure}

We consider the model illustrated in Figure~\ref{fig:system:model}, where clients that are censored and under surveillance need to access sensitive content posted by publishers who are outside the censored area. For this, we define a storage system UWeb = $(\mathcal{B}, ClientSetup, Store, Access)$ built over a blockchain $\mathcal{B}$ that consists of functions to setup client functionality, and store and access content.

Content publishers use the $Store$ function to embed data in the fields of cryptocurrency transactions that are persisted in the blockchain $\mathcal{B}$ at the cost of mining them. Consumers use the $Access$ function to access stored content for free. While UWeb's reliance on cryptocurrency transactions can be leveraged to enable content publishers, content consumers and combinations thereof to pay for the storage of data, the specific payment arrangements are outside the scope of this paper.

Data can be stored in a single transaction or in multiple transactions. We view the blockchain $\mathcal{B}$ to be an ordered set of transactions of the simplified form $(i, d, u)$, where $i$ is the index of the transaction, $d$ is the transaction content, and $u$ is a boolean that specifies whether this is a financial transaction ($u$ = 0) or a data-storing transaction ($u$ = 1).

We assume that any publisher and consumer has control over her computer and can install arbitrary software, including a cryptocurrency node, the cryptocurrency reference software, and the UWeb client that we develop. We assume that developed software can only communicate through the gossip network and the blockchain. While other communication media would simplify the solution design, they often leak sensitive information to the provider~\cite{ISPTracking, ISPSpying}, and can be vulnerable to denial of service and insider attacks. While we assume that software cannot access off-chain storage services, we assume that it can access basic services, e.g., routing, DNS, certification authorities, and other cryptocurrency nodes.

\noindent
{\bf Standard Transactions: Blockchain-Writing Constraints}.
Transactions used by blockchain-writing solutions need to be {\it standard} in order to be relayed through the gossip network, and are eventually mined into the blockchain. For this, they need to satisfy several restrictions imposed by Satoshi blockchains~\cite{isStandardCode}. We now document the most relevant restrictions.

\vspace{-1pt}

\begin{myrestriction} (Size restrictions)
The maximum size of a transaction is 100KB~\cite{maxTxnSize}. The maximum size of a block is 1MB. The individual scripts of an input (output) of a standard transaction cannot exceed a total size of 1,650 KB~\cite{scriptSigActualSize}. Script primitives to operate on data are limited to 512 bytes~\cite{isStandard}.
\label{restriction:size}
\end{myrestriction}

\vspace{-1pt}

\begin{myrestriction} (Input/output count)
Standard transactions can have a theoretical maximum of $2^8$ inputs and outputs. For P2SH, the most common transaction type, the size of a transaction input is 108B and the size of an output is 34B. Assuming one input and given the above maximum transaction size, the maximum number of outputs in a P2SH transaction is 2,937.
\label{restriction:input_output_count}
\end{myrestriction}

\vspace{-5pt}

\begin{myrestriction} (Confirmation time)
Issued transactions need to wait to be confirmed to be mined into a block. Thus, the data-writing process needs to wait one {\it mining epoch} (on the order of minutes) for a segment of data to be confirmed on a block. 
\label{restriction:confirmation_time}
\end{myrestriction}

\vspace{-1pt}

\begin{myrestriction} (Unconfirmed chains)
 The length of a chain of unconfirmed transactions ($\S$~\ref{sec:related}) cannot exceed 25 transactions. The total size of the chain of unconfirmed transactions cannot exceed 101KB~\cite{unconfirmedchains}.
\label{restriction:unconfirmed_chains}
\end{myrestriction}

\vspace{-1pt}

\begin{myrestriction} (Minimum transaction fees) The current minimum relay fee rate is $10^{-8}$ LTC per byte (resp. BTC) for the Litecoin (resp Bitcoin) systems.
\label{restriction:minimum_fees}
\end{myrestriction}

\vspace{-5pt}
\subsection{Adversary Model}
\label{sec:model:adversary}

We consider an adversary who seeks to monitor the access to content of users in a certain region, and even to prevent accesses to content considered sensitive, see Figure~\ref{fig:system:model}. In the following we first detail the monitoring adversary, then the censoring one.

In the following, we say that a network communication $\Gamma$ is possible when accessing transaction $T_i=(i, d_i, u_i)$ if $P(\Gamma|T_i)$ > 0, i.e., the conditional probability that UWeb performed communication $\Gamma$ with the cryptocurrency network to access data $d$ is positive.

We consider adversaries that are able to monitor the communications of UWeb users and other Satoshi nodes, and arbitrarily inject, delete, and reorder messages. We assume however that the adversary cannot decrypt encrypted content or forge signatures, without knowledge of the decryption and signature generation keys, respectively. We assume that the adversary can run any number of clients and services, including of UWeb, and can publish content or access content published by others.

\noindent
{\bf Private Network Access (PNA) Game}.
We say that a communication system that uses blockchain $\mathcal{B}$ provides {\it private access} in terms of network communication, if given a pair of transactions $T_{i_0}, T_{i_1}$ in $\mathcal{B}$, the adversary can not identify the transaction id choice of the UWeb system with probability significantly greater than that of a random guess (1/2). Formally, we want to demonstrate that any probabilistic polynomial time (PPT) adversary $\mathcal{A}$ has only a negligible advantage over random guessing on the following security game against a challenger $\mathcal{C}$ that accesses data stored on $\mathcal{B}$.

\begin{compactitem}

\item
$\mathcal{C}$ installs required software (e.g., Satoshi client) and sets up functionality as described in $\S$~\ref{sec:uweb:uweb}. $\mathcal{A}$ stores the blockchain $\mathcal{B}$ to serve its content over the cryptocurrency p2p network, upon request.

\item
The adversary may perform a polynomially bounded number of operations to select 2 indices $i_0, i_1$ of transactions in blockchain $\mathcal{B}$.

\item
At a desired time, $\mathcal{A}$ sends to $\mathcal{C}$ the indices $i_0$ and $i_1$, where $T_{i_0} = (i_0, d_0, u_0) \in \mathcal{B}$, $T_{i_1} = (i_1, d_1, u_1) \in \mathcal{B}\}$.

\item
$\mathcal{C}$ picks a bit $b \in_R \{0, 1\}$ uniformly at random. $\mathcal{C}$ performs communication $\Gamma$ to access transaction $T_{i_b}$ from $\mathcal{A}$, then signals completion.

\item
$\mathcal{A}$ performs additional computations and outputs bit $b'$, his prediction of the bit $b$ chosen by $\mathcal{C}$.

\end{compactitem}

\noindent
The advantage of $\mathcal{A}$ in this game is:
\[
\begin{aligned}
\textbf{Adv}_{\text{UWeb}}^{\text{PNA}}(\mathcal{A}) &= |P(b' = b) - P(b' \neq b)|
\end{aligned}
\]

We then introduce the following definition:

\begin{definition}(Private Network Access)
A blockchain-based censorship resistant system provides private network access if there exists no PPT adversary with a non-negligible advantage in the PNA game.
\label{def:pna}
\end{definition}

\noindent
{\bf Anonymous Network Access (ANA) Game}.
We say that a communication system that uses blockchain $\mathcal{B}$ provides {\it anonymous access} in terms of network communication, if given a set of transaction ids $[i] = \{i \mid T_i = (i, d_i, u_i) \in B\}$, the adversary can not determine if the consumer is using UWeb. Formally, we want to demonstrate that any probabilistic polynomial time (PPT) adversary $\mathcal{A}$ has only a negligible advantage over random guessing on the following security game against a challenger $\mathcal{C}$ that chooses whether to use UWeb to access data stored on blockchain $\mathcal{B}$.

\begin{compactitem}

\item
$\mathcal{C}$ installs required software (e.g., Satoshi client) and sets up functionality as described in $\S$~\ref{sec:uweb:uweb}. $\mathcal{A}$ stores the blockchain $\mathcal{B}$ to serve its content over the cryptocurrency p2p network protocol upon request.

\item
The adversary may perform a polynomially bounded number of operations to select $m+n>0$ indices $i \in \{1\dots m+n\}$ of transactions $T_i=(i, d_i, u_i)$ in blockchain $\mathcal{B}$ such that $m$ transactions have bit $u$ = 1 and $n$ transactions have bit $u$ = 0.

\item
At a desired time, $\mathcal{A}$ sends to $\mathcal{C}$ the set of $m+n$ indices: $[i] = \{ i \mid i \in 1 ..m+n , T_i = (i, d_i, u_i) \in \mathcal{B}\}$.

\item
$\mathcal{C}$ selects a bit $b \in_R \{0, 1\}$ uniformly at random. $\mathcal{C}$ then performs communication $\Gamma$ to collect all transactions $T_i = (i, d_i, u_i = b)$ from $\mathcal{A}$ (either $m$ or $n$ of them) and signals completion.

\item
$\mathcal{A}$ performs additional computations and eventually outputs $b'$, his prediction of the bit $b$ chosen by $\mathcal{C}$.

\end{compactitem}

\noindent
We define the advantage of adversary $\mathcal{A}$ in the anonymous network access game to be:
\[
\begin{aligned}
\textbf{Adv}_{\text{UWeb}}^{\text{ANA}}(\mathcal{A}) &= |P(b' = b) - P(b' \neq b)|\\
\end{aligned}
\]

We introduce then the following definition:

\begin{definition}(Anonymous Network Access)
A blockchain-based censorship resistant system provides anonymous network access if there exists no PPT adversary who has a non-negligible advantage in the above ANA game.
\label{def:ana}
\end{definition}

\begin{claim}
A blockchain-based censorship resistant system that provides private network access also provides anonymous network access.
\label{claim:pnaimpana}
\end{claim}

\begin{proof}

Let us assume that there exists a PPT adversary $\mathcal{A}$ with advantage $\epsilon$ in the anonymous network access security game (Definition~\ref{def:ana}). Then, we use that adversary against the challenger in the private network access security game (Definition~\ref{def:pna}) to achieve the same advantage $\epsilon$ against that challenger. Formally, we create a new adversary $\mathcal{A'}$ that acts as a challenger with $\mathcal{A}$ in an anonymous access game, and acts as adversary against a challenger $\mathcal{C}$ in a private access game:

\begin{compactitem}

\item
$\mathcal{A}$ sets $m$ = 1 and $n$ = 1, and chooses transactions $T_{i_0}, T_{i_1}$ such that $u_{i_0}=0$ and $u_{i_1}=1$. $\mathcal{A}$ sends $i_0$ and $i_1$ to $\mathcal{A'}$.

\item
$\mathcal{A'}$ forwards them to $\mathcal{C}$, then acts as a proxy between $\mathcal{A}$ and $\mathcal{C}$, enabling $\mathcal{C}$ to fetch blockchain $\mathcal{B}$ from $\mathcal{A}$.

\item
$\mathcal{C}$ picks a bit $b$ randomly then accesses TX $T_{i_b}$.

\item
$\mathcal{A}$ sends to $\mathcal{A'}$ his guess bit $b'$. $\mathcal{A'}$ forwards $b'$ to $\mathcal{C}$.

\end{compactitem}

\noindent
The advantage of $\mathcal{A'}$ in the private network access game is $\textbf{Adv}_{\text{UWeb}}^{\text{PNA}}(\mathcal{A'}) = |P(b' = b) - P(b' \neq b)|$. The advantage of $\mathcal{A}$ in the anonymous network access game with $\mathcal{A'}$ is also $\textbf{Adv}_{\text{UWeb}}^{\text{ANA}}(\mathcal{A}) = |P(b' = b) - P(b' \neq b)|$, which by definition is $\epsilon$. Thus, we have build an adversary $\mathcal{A'}$ that has advantage $\epsilon$ in the private network access game.

\end{proof}

\noindent
{\bf Censoring Adversary}.
We further consider a censoring adversary who attempts to prevent data retrieval from the blockchain for a set of victim participants, e.g., located within a certain geographic region governed by the censor. For this, the adversary will specifically target the communications required to store and retrieve blockchain data.

However, we assume that the censor is not willing to block or significantly hinder cryptocurrency use inside the censored area, due to the associated collateral damage~\cite{holowczak2015cachebrowser, fifield2015blocking}. For instance, China ranks first in the world on the ``activity of non-professional, individual cryptocurrency users, based on how much cryptocurrency they are transacting compared to the wealth of the average person''~\cite{BitcoinAdoption}, while Russia ranks second in the world in trading volume in Bitcoin~\cite{BitcoinTradingVolume}.

We also assume that the censor controls gossip network nodes, mining nodes, and even mining pools. Such a censor can then further launch output and input script modification attacks, a.k.a., sniping and integrity attacks~\cite{SVS17} (see Appendix~\ref{appendix:attacks}), which effectively corrupt data embedded in transactions that have not yet been mined. These attacks duplicate victim transactions, modify the duplicates at will, then rush the fabricated transactions in the p2p network in an effort to have them mined before the original transactions.

However, since a majority attack would damage the trust in the financial aspect of any cryptocurrency, we assume that the censor does not control more than half of the mining power of the network. Recent research~\cite{TSI21} has shown that no country or mining pool controls 51\% of the mining power in Bitcoin. Further, we assume that the censor cannot block the access of honest clients, to all the nodes with access to an honest pool. We now introduce the following definition:

\begin{definition}($p$-Cap Collateral Damage)
We say that a blockchain-based publishing system imposes a $p$-cap collateral damage if the above defined censor who seeks to prevent well-behaved publishers from posting desired content and consumers from accessing published content, will deprive the economy of the censored region of access to a financial market with a capitalization value of $p$.
\label{def:cr}
\end{definition}

\vspace{-15pt}
\subsection{Problem Definitions}
\label{sec:model:problem}

We consider the problem of providing efficient storage of information such that access to the data is hard to monitor or censor. Developed solutions need to satisfy blockchain-imposed restrictions ($\S$~\ref{sec:model:system}) and optimize several relevant metrics, which we now define. We use the term {\it construct} to refer to the solution's basic unit of storing data, e.g., transaction or group of transactions.

\noindent
{\bf Metrics}.
We call the {\it throughput} of the solution to be the size of data made available on the blockchain (gossip) network by the solution per time unit, and {\it goodput} to be the ratio of size of payload, i.e., data embedded in a construct, to the total size of the construct. Finally, we define the {\it cost} of the solution to be the price per byte of stored data, i.e., the total size of the data-storing construct multiplied by the transaction fee, divided by the payload size.

We now define the data storage problem:

\begin{definition}
(Optimal Blockchain Storage)
Given content $C$ of size $N$ and a single funding address $F$, develop a transaction construct and an optimal organization of such constructs that embed content $C$ and are funded from $F$. The constructs should be standard ($\S$~\ref{sec:model:system}), maximize storage throughput and goodput, minimize the storage cost, and be secure against integrity attacks (Appendix~\ref{appendix:attacks}).
\label{def:optimal:storage}
\end{definition}

Further, we define the data access problem:

\begin{definition}
(Private, Censorship Resistant, Efficient Access)
A censored client needs to efficiently access content stored among hundreds of millions of transactions, with private and anonymous network access (Def.~\ref{def:pna} and~\ref{def:ana}) and censorship resilience (Def.~\ref{def:cr}).
\label{def:private:access}
\end{definition}

\vspace{-15pt}
\section{The UWeb System}
\label{sec:uweb}

In this section we first introduce blockchain storage solutions that satisfy Definition~\ref{def:optimal:storage}: techniques to insert data into standard Satoshi transactions, and indexing methods that ensure the scalability of writing large content. Second, we leverage these solutions to build the UWeb storage system defined in $\S$~\ref{sec:model:system} that satisfies Definition~\ref{def:private:access}.

Our system model ($\S$~\ref{sec:model}) assumes publishers on the censor-free region. This enables us to develop constructs that tradeoff detectability for a significant increase over state-of-the-art in the amount of data that can be stored in a single transaction and between mining events.

\begin{figure}[t]
\vspace{-5pt}
\renewcommand{\baselinestretch}{0.5}
\begin{minipage}{0.47\textwidth}
\begin{algorithm}[H]
\begin{tabbing}
XXX\=X\=X\=X\=X\=X\= \kill
				
1.{\mbox{\bf{P2SH.scriptSig(data)}}}\\ 
2.\>{OP\_PUSHDATA2 data.getNext(520)}\\ 
3.\>{OP\_PUSHDATA2 data.getNext(520)}\\
4.\>{OP\_PUSHDATA2 data.getNext(520)}\\
5.\>{OP\_PUSHDATA1 \% push next 4 lines, 79 bytes}\\
6.\>{OP\_PUSH data.getNext(8) OP\_DROP}\\
7.\>{OP\_HASH160 OP\_PUSH data.getNext(20)}\\
\>\>\>\>{OP\_EQUALVERIFY}\\
8.\>{OP\_HASH160 OP\_PUSH data.getNext(20)}\\
\>\>\>\>{OP\_EQUALVERIFY}\\
9.\>{OP\_HASH160 OP\_PUSH data.getNext(20)}\\
\>\>\>\>{OP\_EQUAL}
\vspace{-40pt}
\end{tabbing}
\caption{Smart contract that maximizes data storage using 1,650 B of space available to transaction scripts. The contract pushes 3 chunks of 512 bytes, validated in the redeeming script via hash comparisons. Only the last hash validation pushes a TRUE value to the stack as mandated by the latest standard transaction rules~\cite{isStandard}.} 
\label{fig:fattrans}
\end{algorithm}
\end{minipage}
\end{figure}

\subsection{Max-Size Data Storing Script}
\label{sec:uweb:script}

\vspace{-5pt}

To satisfy Definition~\ref{def:optimal:storage} we first develop input script-writing solutions that achieve optimal script utilization, i.e., maximize input goodput, and simultaneously prevent the input/output modification attacks of $\S$~\ref{sec:model:adversary} and satisfy the blockchain restrictions for standard transactions. We observe that the blockchain restrictions for standard transactions imply that optimal script utilization allows for a maximum of 3 large push operations (1,650 B / 520 B) and a few extra bytes per input script.

Algorithm~\ref{fig:fattrans} shows the proposed smart contract that achieves optimal utilization of the available storage space on an input script under these restrictions and also prevents the transaction integrity attacks described in Appendix~\ref{appendix:attacks}. Similar to Todd's technique~\cite{inputWritingPeterTodd}, we use hash lock constructs (lines 7-9) to protect the data pushed (lines 2-4). However, unlike Todd's technique, we do not sacrifice one of the data push operations (lines 2-4) to include a public key and a signature verification on the redeeming script (lines 5-9). We do not need this verification because our outputs are simple OP\_RET outpoints (right of Figure \ref{fig:basictxn}). Thus, the maximum storage capacity per input script is 1,568 bytes and after accounting for the overhead, the typical size of a funding/spending transaction pair is 1,703 bytes. We analyze the security that this technique provides against input and output modification attacks, in $\S$~\ref{sec:analysis}.

\vspace{-5pt}

\subsection{Data Storage Solutions}
\label{sec:uweb:maxrate}

We now leverage the above data storing script, to introduce {\it max-rate transactions}, the first practical, transaction-based data-storing construct that satisfied Definition~\ref{def:optimal:storage}). Given a single input address with sufficient funds, max-rate transactions achieve a throughput asymptotic to the available bandwidth, entirely through a Satoshi blockchain (gossip) network, at a minimum cost rate, satisfy the blockchain restrictions of $\S$~\ref{sec:model:system}, and prevent the transaction integrity attacks of Appendix~\ref{appendix:attacks}.

\begin{figure}
\centering
\vspace{-5pt}
\includegraphics[width=0.47\textwidth]{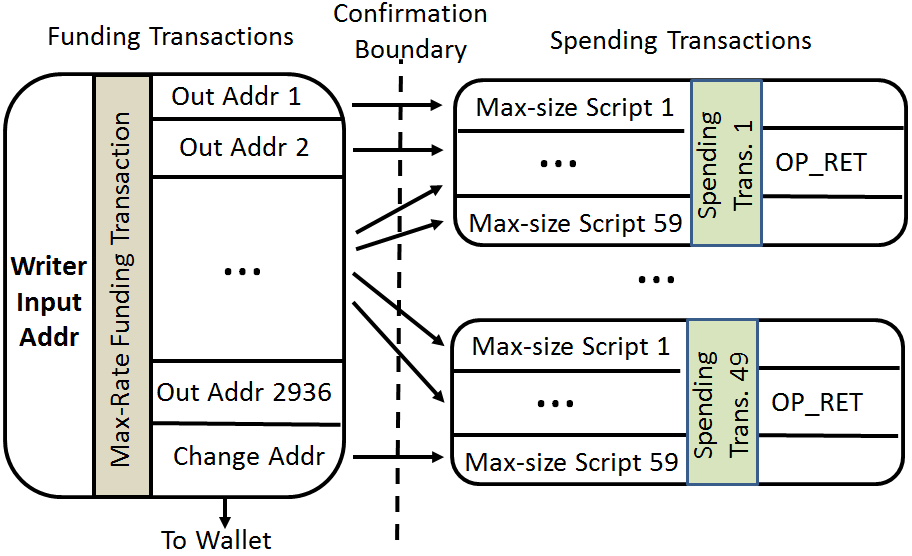}
\vspace{-10pt}
\caption{Optimal data storage construction creates a web of funding and spending transaction that minimizes storage overhead. This basic construct allows for storing maximum file sizes of 4.6 MB in one confirmation epoch.}
\vspace{-15pt}
\label{fig:basictxn}
\end{figure}

To achieve this, we observe that factoring out the transaction overhead, a basic data-storing transaction can handle 59 data storing, spending input scripts (i.e. 100KB / 1,650 B, since the maximum size of a transaction is 100KB) while funding transactions can handle up to 2,937 outputs, with one output reserved for balance change (according to restriction \ref{restriction:input_output_count}, $\S$~\ref{sec:model:system}). Figure~\ref{fig:basictxn}
shows the optimal max-rate transaction construct that allows funding outputs to create spending inputs in a web of transactions that minimizes storage overhead. To optimize storage, we group the maximum amount of inputs into one transaction, thus reduce the overhead of using several transactions for this purpose.

The first issued, {\it max-rate funding transaction} (left of
Figure~\ref{fig:basictxn}) bundles funding outputs Out Addr 1,2,..., $n$, where the maximum $n$ is 2,936 + 1 change output.  Each output is constructed with an address that references the redeeming script of the corresponding payload-storing input (Fat In Script 1,2,..., $n$) inside a {\it max-rate spending transaction} (right of Figure~\ref{fig:basictxn}). As described above, each spending transaction can have a maximum of 59 such payload-storing inputs. A max-rate funding transaction can thus fund 49 (= 2,937 / 59) completely full, fat spending transactions and 1 more spending transaction with 46 inputs. Therefore, the largest amount of data that can be written with this construct (i.e., 1 funding and 50 spending transactions) is 4.6MB (2936*1568).

The max-rate funding transaction and subsequent spending transactions are separated by a confirmation boundary that represents a new block mining event and its respective waiting time. That is, the funding transaction needs to be mined and confirmed before we issue the spending transactions. This is imposed by the blockchain requirement that chains of unconfirmed transactions do not exceed 101KB in size (restriction \ref{restriction:unconfirmed_chains}, $\S$~\ref{sec:model:system}). Thus, every funding output is already confirmed before any spending input enters the network. This ensures minimum wait time for payload storing transactions, since they are no longer constrained by the maximum chain of unconfirmed transactions. We further discuss this in $\S$~\ref{sec:analysis}.

\begin{figure}
\centering
\includegraphics[width=0.47\textwidth]{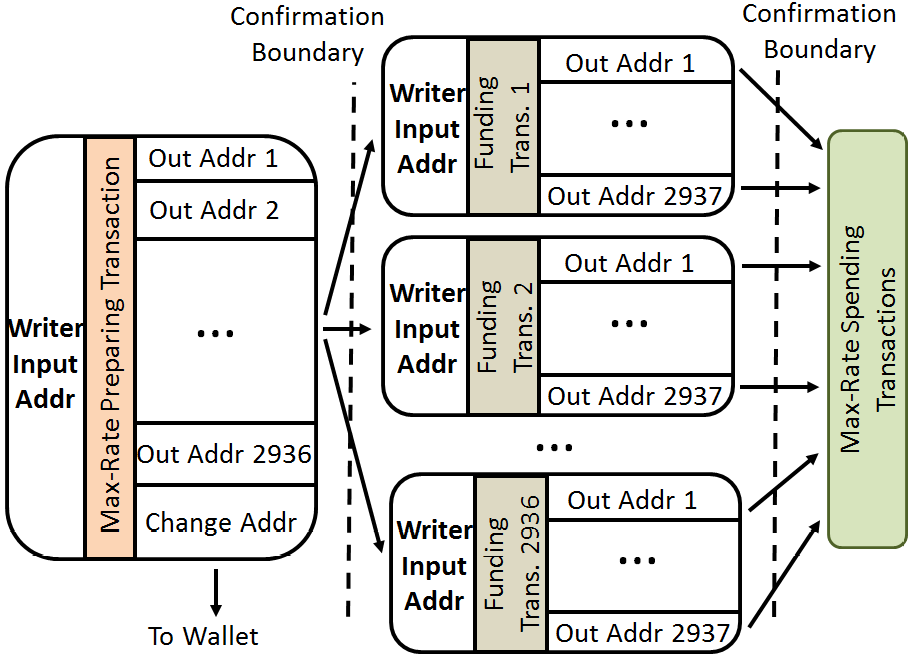}
\caption{Indexing technique that allows the linking of related funding transactions. We store metadata as regular transaction input/outputs in different levels of indirection. The last level consists of the spending transactions that store the actual data (see right side of Figure~\ref{fig:basictxn}). Transactions between consecutive confirmation boundaries, can be sent simultaneously up to the maximum block size, since all their inputs were already confirmed in the previous epoch.}
\vspace{-15pt}
\label{fig:indexing}
\end{figure}

\noindent
{\bf Storing Large Data}. Providing access to large amounts of stored data requires a method for linking together funding transactions that point to related data-storing inputs. The left side of Figure~\ref{fig:indexing} shows the {\it preparing transactions}, a tree-like construct that recursively uses funding transactions to fund other funding transactions. Each level of confirmation boundary multiplies the maximum file size storage by a factor of 2,936: the maximum degree of each \textit{root} node in this tree is 2,936. Thus, approach provides an exponential increase on stored data size, for a linear increase in waiting time.

\vspace{-5pt}

\subsection{UWeb}
\label{sec:uweb:uweb}

\begin{figure}[t!]
\centering
\includegraphics[width=0.99\columnwidth]{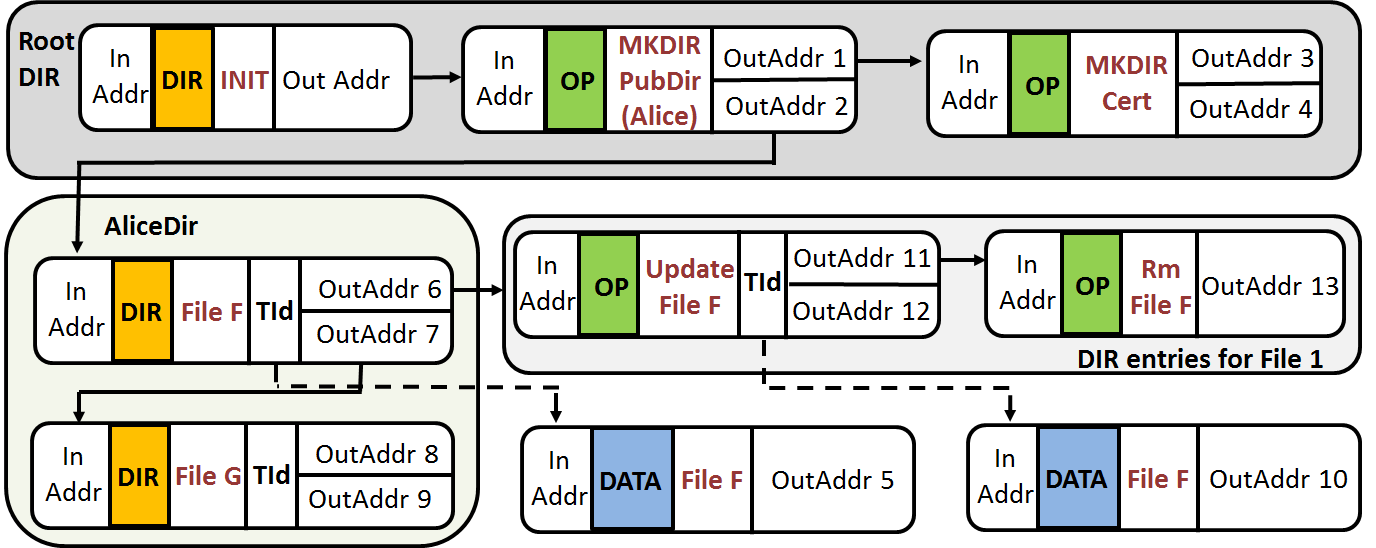}
\caption{UWeb directory organization for publisher Alice. RootDIR points to AliceDIR and pubkey certificate (Cert) subdirs. DIR entries point to either the first DATA transaction that stores content, or to other DIR entry transactions. AliceDir contains two files (F and G). OP entries are chained to record all operations performed on a file (e.g., Update File F, Remove File F).}
\label{fig:dir}
\vspace{-15pt}
\end{figure}

We now leverage max-rate transactions to build the UWeb storage system defined in $\S$~\ref{sec:model:system} and provide private, censorship resistant and efficient access to data stored in a blockchain among millions of transactions. To organize stored content, UWeb defines several content-storing transaction-based entry types, illustrated in Figure~\ref{fig:dir}: {\it DATA entries} are transactions that store compressed content organized in files, and {\it DIR entries} organize information about directories and/or files. DIR entries point to either DATA entries or recursively to other DIR entries. DIR entries also store content meta-data, e.g., whether they point to a file or a directory, and a signature generated by the publisher over the entry's
previous fields, to provide integrity. In the following we define the UWeb functions of $\S$~\ref{sec:model:system}.

\noindent
{\bf ClientSetup Function}.
To set up a content publisher, the $ClientSetup$ function creates a root directory (denoted rootDIR), that contains public identifying information (INIT entries), including, e.g., public key certificates of the publisher,  sub-directories (DIR entries) for other content that include file pointers, publisher metadata and data (DATA entries), e.g., web content.

\noindent
{\bf Content Publishing}.
The $Store$ function takes as input a directory name $Dir$, a file name $Fname$ and content $data$, and uses the DIR and DATA entries defined above to store the content. Specifically, $Store$ first uses the max-rate transaction constructs of $\S$~\ref{sec:uweb:script} to write $data$ to the blockchain, see the DATA-labeled blocks in Figure~\ref{fig:dir}. Let $Tid$ be the Id of the first transaction storing $data$. Then, if an OP entry exists for directory $Dir$ UWeb follows the chain depicted in Figure~\ref{fig:dir} until it finds an un-spent output. Otherwise, it creates a new OP entry for directory $Dir$ and uses one of its outputs. Once such an un-spent output is chosen, the $Store$ function creates a DIR entry with the FILE directive and writes in it the $Fname$ parameter provided as metadata and the above $Tid$, see Figure~\ref{fig:dir}.

All UWeb transaction entries may make use of multiple input addresses in order to ensure the ability to chain these entries even when an output address runs out of enough funding. However, output addresses consist of 1) one OP\_RET output with directives and associated metadata and 2) one or more P2SH outputs for chaining. Because OP\_RET outputs are limited to 80 bytes only, metadata larger than 80 bytes can be stored chaining more DIR INIT entries using the remaining P2SH outputs. In order to mark the end of a metadata chained in this way, we use the same variable length integer format used in Bitcoin \cite{varintformat}. All DIR entries should reserve one P2SH un-spent output in order to continue the chain of directives.

\noindent
{\bf Update: Content Logs}.
The append-only nature of the blockchain makes it challenging to update this directory structure, e.g., to update or remove content. Inspired by log-structured file systems~\cite{RO92, HO95, MMGC02}, the $Store$ function defines {\it OP entries}, to record information about operations performed on files and directories, and appends them to create {\it OP chains}, using Catena~\cite{TD17} chaining. Figure~\ref{fig:dir} illustrates
OP chains, where the first output address of the DIR transaction of a directory/file initiates a chain of all the operations performed on the directory/file. To add to or modify File F (see Figure~\ref{fig:dir}), the $Store$ function first writes the new content (the whole file or only deltas) to a new set of DATA entries, and stores them on the blockchain (see $\S$~\ref{sec:uweb:maxrate}). It then generates an OP entry that embeds the Id of the first transaction that stores this content (i.e., a funding or preparing transaction, see $\S$~\ref{sec:uweb:maxrate}), and funds this transaction from an output address of the transaction that stores the previous
OP entry in the chain (i.e., OutAddr 6, or OutAddr 11 in Figure~\ref{fig:dir}).

\noindent
{\bf UWeb Access}.
The $Access$ function takes as input either a UWeb directory and file name, or a transaction ID. It uses a Satoshi reference client to access the raw underlying cryptocurrency blockchain, for instance~\cite{LitecoinClient}. UWeb has shared access to the internal data storage of the reference client so that its operation is effectively local. As a consequence, the network fingerprint of UWeb content consumers is indistinguishable from a vanilla cryptocurrency user.

To discover content the first $Access$ function call needs to download the entire blockchain. Once the entire blockchain history is available, UWeb scans all transactions looking for DIR entries of different publishers and traverses the structure depicted in Figure \ref{fig:dir}. In order to do this, the client looks for OP\_RET outputs that start with the DIR INIT tag ($0x44495220494E4954$ in hex) and reads off origin metadata including a public key certificate of the publisher. After validating this certificate, downstream content in this data structure can be verified with strong cryptographic assurances.

As UWeb discovers new content, either by traversing the entire blockchain or processing only new transactions, it builds an external database for content random access. When a user searches for content of interest, data is searched in this external database. Thus, accessing content is completely decoupled from the Satoshi reference client operation.

\vspace{-15pt}

\subsection{Analysis}
\label{sec:analysis}

\begin{claim}
The throughput of max-rate transactions is asymptotic to the available network bandwidth.
\label{claim:throughput}
\end{claim}

\begin{proof}
The expected throughput $R(N)$ corresponding to a given size of payload $N$ is given by $R(N) = \frac{N}{w \times E + N/B}$, where $E$ is the number of confirmation epochs (i.e., mining events) required to generate the max-rate transaction, $w$ is the expected wait time for a confirmation event (e.g., $w = 150$ secs for the Litecoin network) and $B$ is the upload bandwidth available to connect to the gossip network. Further, $E = 1+\frac{N}{p \times F}$, where $p$ is the maximum payload size in a script and $F$ is the maximum number of funding outputs that fit in one block. For our max-rate transaction, $p=1,568$ and $F\approx29,370$. For a 46MB payload and a 1Gbs upload bandwidth in our lab, the expected throughput is 154KB/s. Further, the throughput is hyperbolic with respect to the payload size. Thus, when $N\to\infty$, $R(N)\to B$, the available upload bandwidth.
\end{proof}

\begin{claim}
The goodput of max-rate transactions approaches theoretical limit.
\label{claim:goodput}
\end{claim}

\begin{proof}
For content of size $N$, the number of spending transactions is given by $Q(N) = \frac{N}{p \times m}$, where $m$ is the maximum number of max-size inputs ($\S$~\ref{sec:uweb:script}) that fit in a transaction. For Bitcoin/Litecoin, $m = 100$KB$/1,720 \approx 59.5$.  The number of funding transactions is given by $L(N) = \frac{N}{p \times f}$, where $f$ is the maximum number of funding outputs that fit in a transaction. For Bitcoin and Litecoin,
$f=2,937$.  Thus, the total size of max-rate transactions is $S(N) = ts \times \left( \frac{N}{p \times f} + \frac{N}{p \times m} \right)$, where $ts$ is the maximum size of a transaction. For Litecoin, $ts$ = 100KB. Then, the goodput of a max-rate transaction is $G(N) = \frac{N}{S(N)}$.
\end{proof}

\begin{claim}
Max-rate transactions are standard.
\label{claim:standard}
\end{claim}

We include the proof in Appendix~\ref{appendix:proofs}. Further, in Appendix~\ref{appendix:proofs} we also show that max-rate transactions prevent input and output modification attacks. Then, given that our blockchain-writing constructs are standard, maximize throughput and goodput and are secure against input and output modification attacks, we conclude that max-rate transactions satisfy the optimal blockchain storage problem (Def.~\ref{def:optimal:storage}).

\begin{claim}
UWeb provides private network access.
\label{claim:pna}
\end{claim}

\begin{proof}
Following the private network access game (Definition~\ref{def:pna}), UWeb first downloads the entire blockchain before accessing all data-containing transaction $T_i$ (see $\S$~\ref{sec:uweb:uweb}). Thus, the only possible communication $\Gamma$ is the entire log of transactions in blockchain $\mathcal{B}$ until all $T_i$ have been observed. Therefore, $P(\Gamma = \mathcal{B} \mid b) = P(\Gamma = \mathcal{B}) = 1$ for all of $\mathcal{C}$'s choices of $b$, i.e., no matter the choice of $b$, $\mathcal{A}$ always observes the same communication. This implies that knowledge of $\Gamma$ does not provide $\mathcal{A}$ with an advantage in the choice of $b$. Formally, for any PPT $\mathcal{A}$, the conditional probability of guessing $\mathcal{C}$'s $b$ value given communication $\Gamma$ is, according to Bayes' theorem, $P(b' = b \mid \Gamma = \mathcal{B}) = P(\Gamma = \mathcal{B} \mid b) \times P(b)/ P(\Gamma = \mathcal{B})$. Since the choice of $b$ is random (see the PNA game in $\S$~\ref{sec:model:adversary}), $P(b) = 1/2$, thus $P(b' = b \mid \Gamma = \mathcal{B}) = 1 \times (1/2) / 1 = 1/2$. Thus, we have that for any adversary $\mathcal{A}$:
\[
\begin{aligned}
\textbf{Adv}_{UWeb}^{PNA}(\mathcal{A}) &= |P(b' = b) - P(b' \neq b)|\\
& = |P(b' = b) - (1 - P(b' = b))|\\
& = |2 \times P(b' = b \mid \Gamma = \mathcal{B}) - 1| = 0
\end{aligned}
\]
\end{proof}

UWeb also provides anonymous network access: due to Claim~\ref{claim:pnaimpana}, since UWeb provides private network access it also provides anonymous network access.

\noindent
{\bf UWeb provides $p$-censorship resilience}.
First, we observe that since max-rate transactions are standard (see Claim~\ref{claim:standard}), nodes following the protocol will broadcast them in the p2p network and the consensus protocol will eventually add them permanently to the blockchain. Further, a censor cannot use input and output attacks to prevent a publisher outside the censored area from using UWeb to distribute information and persist it in the blockchain, see above discussion.

A censor who seeks to block access to the whole blockchain will trivially deprive the entire censored area of access to the market of the corresponding cryptocurrency. A censor who seeks to selectively filter certain blocks, e.g., containing transactions that embed UWeb content, would generate a split universe with global impact on the consensus protocol, thus will eventually similarly deprive the entire censored area of access to the market of the corresponding cryptocurrency. Given that UWeb can be ported to all Satoshi compliant cryptocurrencies, a censor that wants to block UWeb needs to block all Satoshi compliant cryptocurrencies, thus depriving its economy from access to a market capitalization of p = \$1 trillion (see Definition~\ref{def:cr}).

\vspace{-15pt}

\section{Evaluation}
\label{sec:evaluation}

In this section we experimentally evaluate max-rate transactions. We first detail our experimental setup, then reveal details of a longitudinal storage of BBC news and a stress-test of the Litecoin blockchain. We performed our experiments on the Litecoin blockchain, that required thus real money. All script constructs proposed in this paper were accepted by their networks without any modification, as expected for standard transactions. 

In the following we use two more metrics, in addition to the ones introduced in $\S$~\ref{sec:model:system}. First, the {\it block space utilization} of a blockchain-writing solution, defined to be the ratio of the payload size in a block, to the total size of the block. Second, the {\it block transaction utilization} of a blockchain-writing solution, defined to be the ratio of the number of payload carrying transactions to the total number of transactions in a block.

\vspace{-5pt}

\subsection{Experimental Setup}

\noindent
{\bf The Max-Rate Toolset}.
We implemented a suite of python scripts (1,765 loc) for creating, sending, reading, and monitoring max-rate transactions, that avoided using Litecoin library dependencies as we required the creation of highly customized smart contract constructs. We created a different set of python and bash scripts (118 loc) for monitoring and analyzing mempool statistics, and for scraping and processing journalistic data for its eventual publishing.

\noindent
{\bf Instrumenting the Litecoin Network}.
In order to evaluate the impact of max-rate transactions over normal blockchain operations, we instrumented the Litecoin network to measure the size of the Litecoin mempool and the expected time to confirmation for new unconfirmed transactions. Specifically, we created a script that queried the mempool of a node in our lab every 5 seconds. After each query, we recorded timestamps, size, fee and block height for newly received transactions. Further, every 10 minutes, another script scanned the collected list looking for transactions with 1 confirmation, and calculated the time required for them to make it to the blockchain.

\noindent
{\bf Equipment}.
For our experiments we used two virtual machines with 8 core Intel(R) Xeon(R) Gold 6126 CPU @ 2.60GHz, 8 GB of RAM and 500 GB of hard drive each. We also deployed our scripts on a Raspberry PI 3B+ with an ARM Cortex-A53 CPU @ 1.4GHz, 1 GB of RAM and 32 GB SDHC card, and the Raspbian OS. We downloaded the entire Litecoin blockchain on the PI and validated the execution of the max-rate toolsuite for payloads up to 40 MBs.

\vspace{-5pt}

\subsection{Ethical Considerations}

\noindent
{\bf UWeb Impact on Full Nodes}.
UWeb provides privacy access assurance for arbitrary data inserted in the blockchain. This may suggest that this will inevitably lead to an excessive bloat of the blockchain, and will act as a counterincentive for full nodes to continue to participate. We note however that UWeb imposes costs on writing efforts, thus max-rate transactions are equivalent to a surge in popularity of financial transactions with a minimum transaction fee rate. Dealing with such an event was considered by the Satoshi design; the following experiments show that it works as expected. Thus, this problem is within the scope of the natural evolution of cryptocurrency systems.

\noindent
{\bf UWeb Impact on Financial Transactions}.
Our experiments in the Litecoin realnet ($\S$~\ref{sec:evaluation:bulk}) and simulations ($\S$~\ref{sec:evaluation:simulation}) reveal that UWeb imposes no delays on the confirmation times of financial transactions.

\noindent
{\bf Objectionable Content}.
UWeb may seem to enable the distribution of objectionable content in an unbounded fashion throughout the blockchain. We note however that UWeb is not the ideal tool for posting objectionable content since it does not provide writers with privacy or anonymity (UWeb provides access privacy/anonymity only for readers). Objectionable content already exists in blockchains, and was written with other, much less efficient but more private writing solutions, e.g., overwriting the transaction value field ~\cite{bitcoinBotnets}.

We note that promising research on \textit{redactable blockchains} has been conducted by Deuber et al.~\cite{DMK2019}, and also~\cite{politou2019blockchain,florian2019erasing,derler2019fine}. We envision a system where it is cheaper for a majority to delete undesirable content than for a minority to keep re-introducing it. Such a system would self-regulate the impact of arbitrary data insertions on the overall blockchain ecosystem. We acknowledge however that (1) content could be written encrypted, for valid confidentiality reasons, and (2) mining pools and regular nodes need incentives to correctly implement such blockchain redactions. \newcontent{Further, we observe that un-checked blockchain redactions are a threat to UWeb's censorship-resistance.}

\vspace{-15pt}

\subsection{Longitudinal News Feed Writing}
\label{sec:evaluation:bbc}

\begin{figure}[t!]
\centering
\includegraphics[width=0.47\textwidth]{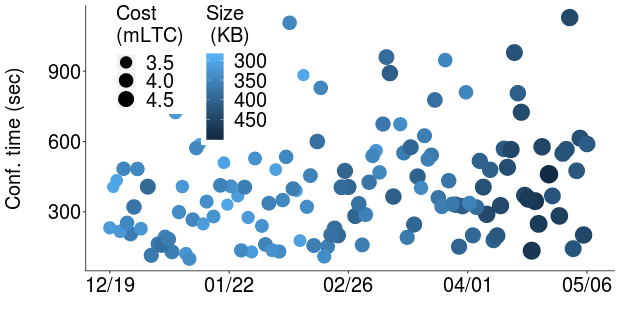}
\caption{Writing BBC articles daily on the Litecoin blockchain. The cost of	writing does not impact the time to first confirmation.}
\label{fig:time_cost}
\vspace{-15pt}
\end{figure}

We obtained permission from the British Broadcasting Corporation (BBC) to reproduce their web articles (only text) into the Litecoin blockchain. We created a crawler that de-fanged articles (also removed the ads) from all public BBC's RSS feeds every day for 134 consecutive days. We daily bundled the text-only articles (average of \the\numexpr (\totalarticles/\entries)~daily articles, at an average of \the\numexpr (\size/\totalarticles)~ characters and 555 words per article) into categorized directories, compressed them into a gzipped file and wrote the archive using max-rate transactions ($\S$\ref{sec:uweb:maxrate}).

In total, we performed \entries ~writing operations (1 per day) to store \numprint{\size} ~Bytes for \numprint{\totalarticles} ~articles with an accumulated cost of \roundfour{\cost}\result ~LTC (\$\ltcToUSD{\cost}\result ~at the time of writing). Thus, the average cost per article is \ltcToLito{\avgcost}\result ~Litoshi (\$\ltcToUSD{\avgcost}\result ~at the time of writing), and the average daily cost is 0.0041 LTC (\$\ltcToUSD{0.0041}\result). Assuming a daily average of 400 articles, at an average of 1000 characters per article, we expect the daily storage size to be around 400 KB. At this publication rate and current LTC price, we expect a monthly publication cost of \$6 USD, which is below the average monthly news digital subscription price in the US of \$9.24~\cite{publishingcost}.

\begin{table}
\centering
\resizebox{0.47\textwidth}{!}{
\textsf{
\begin{tabular}{lcccccc}
\toprule
& Min  & 1st Qu. & Median & Mean & 3rd Qu. & Max. \\
\toprule
Daily articles & 340.0 &  380.0 &  394.0 &  401.6 &  421.8  & 483.0   \\
Size (Kbytes) & 302.5  & 341.9  & 361.4  & 370.7 &  397.5 &  495.8 \\
TX time (sec) & 31.28 &  78.31  & 142.97 & 190.48 & 259.38 & 951.28 \\
Conf. time (sec) & 99.82 & 241.34 & 368.06 & 412.21 & 532.84 & 1128.39 \\
Cost (mLTC)  & 3.478 &  3.931  & 4.155  & 4.263  & 4.573 &  5.700 \\
\bottomrule
\end{tabular}}}
\caption{Statistics from UWeb writing of BBC articles.}
\label{tables:bbc:stats}
\vspace{-25pt}
\end{table}

Each daily storage operation required one funding transaction and four spending transactions. Each storing operation achieved an average transmission throughput of \roundtwo{\avgswift}\usd{\result} Bytes/sec and an average confirmation throughput of \roundtwo{\avgonchain}\usd{\result} Bytes/sec. Figure \ref{fig:time_cost} shows the daily time required for all transactions to obtain at least one confirmation on the blockchain. This time corresponds to the confirmation time row in Table~\ref{tables:bbc:stats}. The color intensity of each dot represents the size in bytes of the bundle of daily articles, whose statistics are shown in the size row of Table~\ref{tables:bbc:stats}. Further, the dot radius in Figure~\ref{fig:time_cost} represents the cost in mLTC, which corresponds to the cost row in Table \ref{tables:bbc:stats}. These results suggest that the transaction cost and size do not have a consistent impact over
the expected confirmation time. On the contrary, confirmation time may be more sensitive to block utilization and network congestion due to large mempool sizes, as we shall see in the following.

\begin{figure}[t!]
\centering
\includegraphics[width=0.47\textwidth]{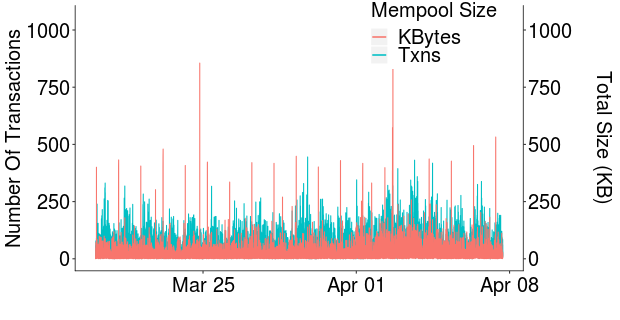}
\vspace{-10pt}
\caption{Timeline of mempool size, with 1 min granularity, of (1) total size in KB, $y$ axis on the right, and (2) number of unconfirmed transactions, $y$ axis on the left. BBC article writing activity is responsible for the regular red spikes in total mempool size. Blue spikes are un-correlated to the red spikes, suggesting that temporary increases in activity do not pose a risk to the current network operation.}
\label{fig:mempool_size}
\vspace{-5pt}
\end{figure}

\noindent
{\bf Impact on Litecoin Network}.
Figure~\ref{fig:mempool_size} shows the network mempool size evolution during a subset of 18 days of the BBC writing experiment, between March 20, 2019 and April 7, 2019. We observe that the BBC writing operations generate substantial, periodic spikes in terms of mempool size (in KB).  However, these spikes quickly subside. We attribute this to the efficient blockchain utilization of max-rate transactions: all BBC writing transactions are likely embedded in the same block, which, when mined, frees mempool space. We note however that since each BBC writing operation fits in only 5 transactions, its impact on the number of unconfirmed transactions in the mempool is negligible. In addition, we observe that blue (transaction count) spikes are not correlated to the BBC writing red (total size) spikes. This suggests that max-rate transactions achieve efficient block utilization, by occupying large portions of the block size while using a few transactions.

\begin{figure}
\centering
\includegraphics[width=0.47\textwidth]{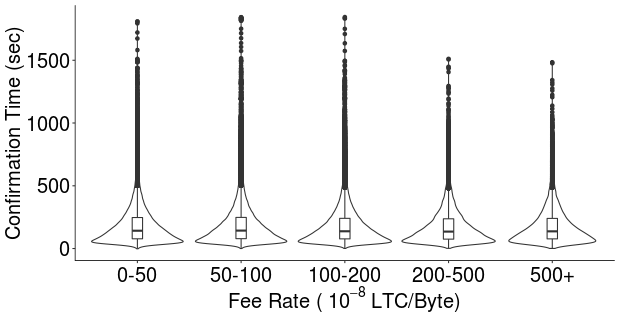}
\vspace{-10pt}
\caption{Violin plots of new transactions confirmation time versus employed fee rate. Small bursts of arbitrary (BBC) writing do not have a measurable impact over transaction confirmation times.}
\label{fig:txn_conf_time}
\vspace{-15pt}
\end{figure}

Further, we evaluated the impact of the BBC writing operations, on the time it takes Litecoin transactions of other users to be confirmed in the blockchain. Figure~\ref{fig:txn_conf_time} shows the time required to achieve one confirmation for different fee rate levels during the March 20 - April 7 interval, over all the transactions posted on Litecoin during that interval. The time required to confirm new transactions is consistent with the expected 2.5 min epoch time.  Moreover, for these levels of block utilization, the fee rate does not appear to have any significant impact over the network throughput.

\vspace{-15pt}

\subsection{Blockchain Stress-Test}
\label{sec:evaluation:bulk}

\begin{figure}
\centering
\includegraphics[width=0.47\textwidth]{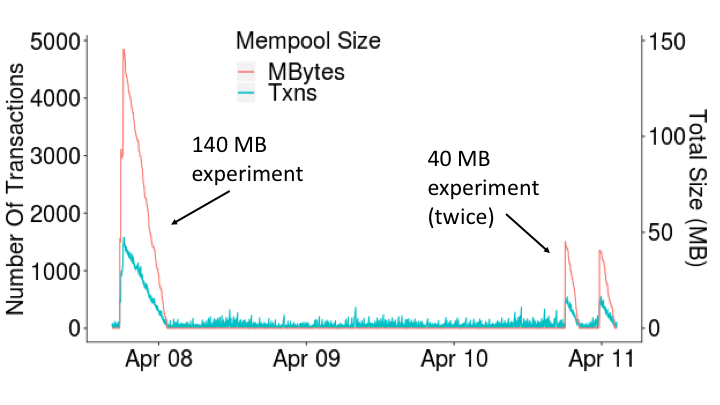}
\vspace{-10pt}
\caption{Mempool size in total number of bytes and number of unconfirmed transactions, during our experiments.}
\label{fig:mempool_size_exp}
\vspace{-5pt}
\end{figure}

To evaluate UWeb and max-rate transactions under large writing loads, we have collected a set of 41 censorship resistant (CR) tools from~\cite{crtools} that include Tor\cite{getTor}, Stegotorus\cite{stegotorus}, Uproxy-client\cite{uproxyClient}, Bit-smuggler\cite{bitsmuggler} and Shadowsocks. We excluded projects that did not publish an explicit re-distribution license. Table~\ref{tables:tfs:licenses} in Appendix~\ref{appendix:crslicenses} provides the full list. The total compressed size of the 41 tools is 217 MB.

We designed our experiment in two phases. First, we almost-concurrently sent the first 3 largest size projects as individual archives, for a total of 140 MB. Second, we sent the rest of the tools compressed, in one split archive, for a total of 77 MB.  This strategy allowed us to send files of size at most 45MB each, compatible with our virtual machine RAM memory resources. The total combined cost for this experiment was 2.51315448 LTC (\$\ltcToUSD{2.51315448}\roundtwo{\result}\result ~at the time of writing).

\noindent
{\bf Mempool and Confirmation Time Impact}.
Figure~\ref{fig:mempool_size_exp} shows the timeline of the mempool size in terms of number of transactions ($y$ axis left side) and total byte size ($y$ axis right side), during the 3 writing operations. We observe that our writing experiments generate substantial spikes in terms of both number of transactions and their total size, which last for 7.65, 2.62 and 2.36 hours for the 140MB, 41.9MB and 38.5MB writing experiments, respectively. Further, the mempool is vacated linearly, which implies that higher fee transactions get prioritized correctly and our actions do not produce a degradation at current network usage levels.

\begin{figure}
\centering
\includegraphics[width=0.47\textwidth]{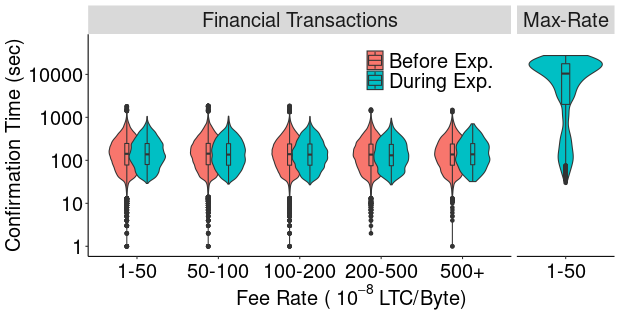}
\vspace{-10pt}
\caption{Violin plots of confirmation times during and before our experiment. Financial transactions on the left, max-rate transactions on the right. Large-scale writing in the blockchain has no impact on financial transactions of any fee rate. Max-rate transactions however experience significant delays, up to 27,500s.}
\label{fig:onchain_time_composed}
\vspace{-15pt}
\end{figure}

We then measured the time each new unconfirmed transaction spent in the mempool before obtaining one confirmation on the Litecoin blockchain. Figure~\ref{fig:onchain_time_composed} compares the distribution of the confirmation time ($y$ axis, log scale) over transactions employing different fee rates ($x$ axis), during (1) our 140 MB writing experiment that lasted 7.6 hours from the first unconfirmed max-rate transaction observed in the mempool, until the last one received at least one confirmation (blue violins), and (2) before our experiment: since the start of the mempool instrumentation measurements on March 20, 2019, until the first unconfirmed transaction of the 140 MB experiment arrived to the mempool on April 7, 2019 (red violins). Interval (2) is the same 18 days when we monitored the mempool during the BBC experiment.

\begin{figure*}
\centering
\includegraphics[width=0.97\textwidth]{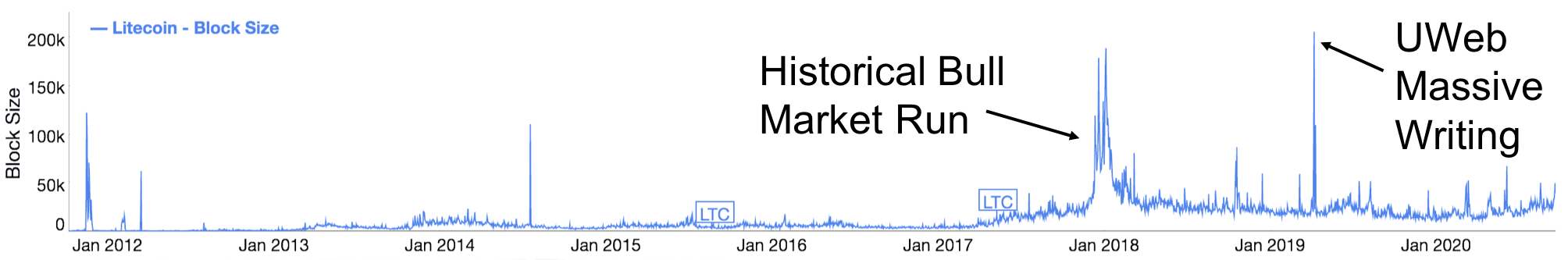}
\caption{Bulk writing experiment achieved the largest daily average block size (206.02KB) ever recorded on the history of Litecoin, exceeding the former record of the bull market run of 2017-2018 (189.1KB).}
\label{fig:bitinfocharts}
\vspace{-5pt}
\end{figure*}

We observe that our significant writing experiment did not impact the confirmation times of financial transactions, including those employing the lowest fee rates (1- 50 lit/B). However, the maximum max-rate transaction confirmation time was 7.63 hours.

Financial transactions are unaffected because mining pools greedily select the next block to mine based on the fee rate of the transactions in the mempool~\cite{BlockSelection}. Max-rate transactions have the lowest fee rate (1 litoshi per byte) thus have the lowest priority for selection. \newcontent{Only four financial transactions (of sizes up to 1,994B) had a fee rate of 1 litoshi/B. Their delays ranged from 69s to 527s (8.8 mins). This suggests that an overwhelming majority of financial transactions use fee rates that exceed the UWeb fee rate, thus are not impacted. While it is theoretically possible for minimum fee-rate transactions to be impacted by UWeb, this did not occur in our experiments.}

\begin{figure}
\centering
\includegraphics[width=0.47\textwidth]{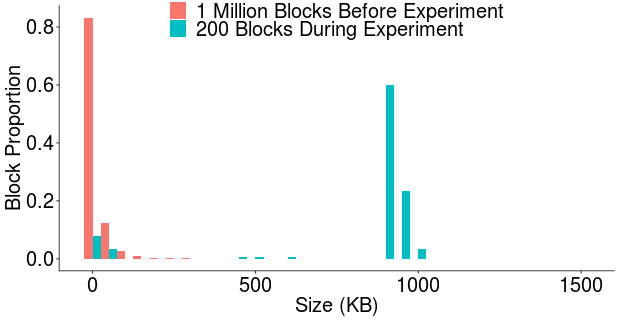}
\vspace{-10pt}
\caption{Block utilization comparison before and during writing  experiment. The distribution of block utilization is preserved even for very large periods of historical data.}
\label{fig:block_utilization_cpm}
\vspace{-10pt}
\end{figure}

\noindent
{\bf Goodput and Throughput}.
We evaluate the goodput and throughput achieved during the first large-scale writing experiment (140+ MB data, over 7.6 hours) on the Litecoin blockchain. We split the aggregated data into 3 chunks of around 45 MBs each and sent them sequentially near in time. We took this precaution to have more control over deciding whether to call off the experiment in case the network was unable to handle the load. The average goodput measured over these 3 combined experiments was 90.8\%, consistent with our theoretical analysis in $\S$~\ref{sec:analysis}. The average throughput measured over these experiments, assuming 3 available funding inputs, was 183 KB/s. This exceeds the theoretical throughput of $\S$~\ref{sec:analysis}, that considers a single funding input.

\noindent
{\bf Block Utilization}.
Figure~\ref{fig:block_utilization_cpm} compares the block size distribution over 200 blocks mined during the first large-scale writing experiment (blue bars) and over 1 million blocks mined before the experiment. We observe that most blocks mined during our experiment are close to the 1MB block-size limit, in contrast to the small size of blocks mined before.

Our large-scale writing experiments achieved a lifetime record high daily average block size as recorded by bitinfocharts~\cite{bitinfocharts}. Figure~\ref{fig:bitinfocharts} shows the daily average block size of the Litecoin network for its entire lifetime. During the day of our experiments, the average daily block size reached 206KB, the largest ever recorded in the history of Litecoin.

In Appendix~\ref{appendix:tor} we zoom-in into block utilization results when using UWeb to write the Tor source code~\cite{torsource} compressed archive (6.4 MB).

\vspace{-5pt}

\subsection{Max-Rate Transactions vs. The World}
\label{sec:evaluation:state}

\begin{table*}
\centering
\resizebox{\textwidth}{!}{
\textsf{
\begin{tabular}{lcccccccccc}
\toprule
Insertion  & Cost/Byte  & Max payload/ & Construct & Construct&Block Space&Block Txn & Safe to output & Safe to input & Private & Censorship\\
Technique & (Sat-Lit) &  Construct & Goodput (\%) & Throughput&Utilization (\%) &Utilization (\%) & mod. attack & mod. attack & access & Resilience\\
\toprule
Apertus & 787.3 & 20 B &13.1 & 0.3 B/min & 0.016&0.052 & \Large \color{ForestGreen}\checkmark & \Large \color{ForestGreen}\checkmark & \Large \color{Red}$\times$ & \Large \color{ForestGreen} \checkmark\\
Catena & 205.62 & 80 B & 34.0 & 1.3 B/min & 0.026&0.052 &  \Large \color{ForestGreen} \checkmark & \Large \color{ForestGreen}\checkmark & \Large \color{Red}$\times$ & \Large \color{ForestGreen} \checkmark\\
MoneyMorph(Bitcoin) & 359.63 & 20 B & 13.1 & 0.3 B/min & 0.016&0.052 &  \Large \color{ForestGreen}\checkmark & \Large \color{ForestGreen}\checkmark & \Large \color{ForestGreen}\checkmark & \Large \color{ForestGreen} \checkmark\\
Tithonus & 1.12 & 1635 B & 88.8 & 545 B/s & 0.20&0.052 & \Large \color{Red}$\times$ &\Large \color{Red} $\times$ & \Large \color{ForestGreen} \checkmark & \Large \color{ForestGreen} \checkmark\\
{\bf Max Trans.} & 1.12 & {\bf 46.3 MB} & {\bf 90.8} & {\bf 115.1 KB/s} & {\bf 88} & {\bf 30} & \Large \bf \color{ForestGreen}\checkmark & \Large \bf \color{ForestGreen}\checkmark & \Large \bf \color{ForestGreen} \checkmark & \Large \bf \color{ForestGreen} \checkmark\\
\bottomrule
\end{tabular}}}
\begin{flushleft}
\end{flushleft}
\vspace{-20pt}
\caption{Comparison of blockchain-writing techniques, considering the availability of a single spending address. In our experiments, the max-rate transaction approach (last row) achieves 2-3 orders of magnitude improvements in throughput and block utilization compared to state-of-the-art solutions, and a goodput improvement of 7 percentage points, while also providing private access, censorship resilience, and resilience to transaction integrity attacks.
}
\label{tables:solutions:comparison}
\vspace{-15pt}
\end{table*}

Table~\ref{tables:solutions:comparison} compares max-rate transactions against state-of-the-art blockchain-writing solutions, on the metrics included in Definition~\ref{def:optimal:storage}.  In our experiments, max-rate transactions achieve 2-3 orders of magnitude improvements in throughput and block utilizations and are the only solution that provides private access, censorship resistance, and resilience to integrity attacks. Catena~\cite{TD17} and Apertus~\cite{Apertus} embed tens of bytes per input in a transaction, and have confirmation time requirements. This, coupled with the required transaction fees and the fact that often these transactions are non-redeemable, also leads to impractical costs for writers.

While Tithonus~\cite{tithonus} has the same cost and slightly lower goodput to max-rate transactions, UWeb significantly improves on its throughput. Since UWeb does not attempt to publish content from within the censored region, it significantly improves on the throughput, goodput and cost efficiency of MoneyMorph~\cite{MMK20}.

Further, we notice that the goodput of max-rate transactions is consistent with the theoretical value derived in the proof of Claim~\ref{claim:goodput}. The throughput achieved by max-rate transactions is also consistent with the theoretical value derived in the proof of Claim~\ref{claim:throughput}. In contrast, the constructs used by Apertus~\cite{Apertus}, Catena~\cite{TD17} and Blockstack~\cite{ANSF16} have a constant relationship between payload size and throughput, thus cannot reduce the ratio between the maximum and their observed empirical throughput.

\vspace{-5pt}

\subsection{Simulation Stress-Test}
\label{sec:evaluation:simulation}

\begin{figure}
\centering
\includegraphics[width=0.47\textwidth]{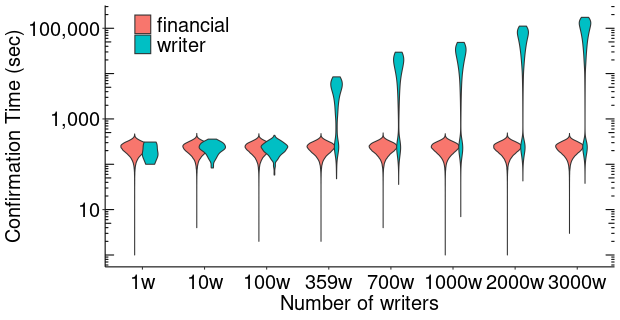}
\vspace{-10pt}
\caption{Violin plots of confirmation times (y axis log scale) of financial and max-rate transactions when up to 3,000 BBC writers are active concurrently. Financial transactions are not affected by UWeb writing. This confirms our realnet experiments. Once the 1MB block size is reached, the confirmation time of max-rate transactions increases linearly with the number of writers.}
\label{fig:sim:writer:delays}
\vspace{-15pt}
\end{figure}

We performed simulations to understand the impact on the Litecoin blockchain of multiple concurrent UWeb writers and higher levels of financial transactions. The simulator generates the next block to be mined by prioritizing the transactions from the mempool that have the highest fee rate~\cite{BlockSelection}, and breaks the tie based on arrival times.

\noindent
{\bf Concurrent Writers}.
To evaluate UWeb and the Litecoin ecosystem under various loads of concurrent writers, we performed a simulation using the trace of financial transactions recorded in the Litecoin blockchain between April 7 and April 10. This interval includes the interval of the realnet experiment shown in Figure~\ref{fig:onchain_time_composed}.

We have simulated between 1 and 3,000 concurrent writers, each writing 400KB in the same 4 hour interval. Thus, each writer is equivalent to the daily writing load from the BBC experiment ($\S$~\ref{sec:evaluation:bbc}). We have added max-rate transactions from each writer (four of size 99,931B one of 10,876B) at times randomly distributed in the same four-hour interval at the beginning the financial transaction trace. The case of 359 writers is roughly equivalent to the realnet experiment whose results are shown in Figure~\ref{fig:onchain_time_composed}, where we wrote 140MB in the Litecoin blockchain. The 3,000 writers post 1.14GB of data-storing transactions.

Figure~\ref{fig:sim:writer:delays} shows the distributions of confirmation times of financial and max-rate transactions. We observe that financial transactions (red violins) are not impacted by the number of writers. This is because all financial transactions have a fee rate that exceeds UWeb's 1 litoshi/B rate. The maxrate transactions of UWeb writers have similar confirmation times to financial transactions, up to a number of writers that do not consume the available space in a block. After that point the confirmation time for maxrate transaction increases linearly with the number of writers.

\begin{figure}
\centering
\includegraphics[width=0.47\textwidth]{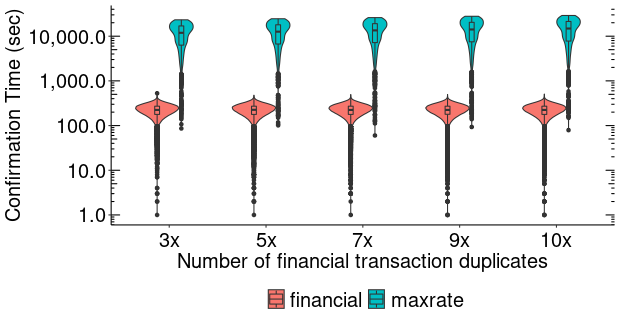}
\vspace{-10pt}
\caption{Violin plots of confirmation times (y axis log scale) for the experiment depicted in Figure~\ref{fig:onchain_time_composed}, when the number of financial transactions is up to 10 times the level in the experiment. Financial transactions are not impacted, while max-rate transactions experience a linear increase.}
\label{fig:sim:financial:delays}
\vspace{-15pt}
\end{figure}

\noindent
{\bf Higher Levels of Financial Transactions}.
We have further simulated the impact of higher levels of financial transactions on the 140MB realnet-writing experiment depicted in Figure~\ref{fig:onchain_time_composed}. For this, we have generated synthetic transaction traces where each financial transaction in a 36 hour interval following the start of the data writing experiment, contributes between 3 to 10 copies. Each copy of a financial transaction arrives at a random time during the 36h experiment. Figure~\ref{fig:sim:financial:delays} shows the distributions of the confirmation times. We observe that financial transactions are not impacted by this increased load, experiencing a mean of around 224s and maximum of 527s in all experiments. However, the confirmation times of maxrate transactions experience a linear increase with the density of financial transactions. For instance, at 3x financial transactions, the mean confirmation delay for max-rate transactions is 11,646s (SD = 6508, max = 23,253), while at 10x the mean delay is 14,546 (SD = 8,213, max = 29,101).

\noindent
{\bf Blockchain Writing Throughput}.
The per-writer blockchain throughput is upper bounded by $A/150 n$, where $A$ is the available block space (1MB minus the financial transactions from the mempool), 150s is the average epoch length, and $n$ is the number of writers. We note however that UWeb transactions will reach all the online UWeb readers in a few seconds over the Litecoin p2p gossip network~\cite{RC19}. Offline UWeb readers are not impacted by high confirmation delays: once a reader is back online they can either access newly written transactions in the mempool or in the blockchain: during the 3,000 concurrent writers experiment, the mempool had at its peak 11,379 max-rate transactions with a total of 1.05GB, and all these transactions received at least one confirmation within 52.27 hours.

\vspace{-5pt}

\section{Discussion}
\label{sec:discussion}

\vspace{-5pt}

\noindent
{\bf What If China Censors Blockchains?}
China has not blocked cryptocurrency-carrying traffic for almost 13 years despite earlier reports of their use to evade its firewall~\cite{R19, Z18, K19}. Further, UWeb is compatible and can be ported to all Satoshi blockchains. Thus, a censor that would block only a subset of Satoshi blockchains would still be unable to prevent UWeb use. A censor that blocks all Satoshi cryptocurrencies will deprive the economy of the censored region of access to a financial instrument with a cap that exceeds \$1 trillion at the time of writing. We also note that UWeb provides value for all countries with restricted freedom of speech \cite{RSFRankings} that do not block Satoshi cryptocurrencies.

\noindent
{\bf Why Not Other Cryptocurrencies?}
Other cryptocurrencies can also provide the underlying censorship-resistant storage medium for UWeb, particularly Monero and Zcash that offer additional privacy guarantees. MoneyMorph~\cite{MMK20} compared Monero, Zcash, Bitcoin, and Ethereum, and revealed that Zcash shielded addresses offer the highest censorship-resistant bandwidth by allowing the embedding of up to 1,148B. Other efforts have also used the Ethereum blockchain to avoid censorship resistance~\cite{K19, Z18}.

Our choice of Satoshi cryptocurrencies is based on their huge share of the cryptocurrency market cap (62\%)~\cite{coinmarketcap}, a network of 10,000 P2P nodes~\cite{bitnodesBitcoinNodesDistribution} and a large and growing hashrate (150m Th/s)~\cite{blockchaincomBitcoinHashrate}.

In this paper we also developed solutions that embed up to 1,650B in a Satoshi address and up to 46.3MB of data in a staging and spending transaction construct. Our solutions achieve a writing throughput that exceeds those of state-of-the-art solutions by 2-3 orders of magnitude. We hope that our work will encourage further research in the use of Monero, Zcash and Ethereum as a medium for censorship resistance.

We also note that our main reason for conducting realnet experiments on Litecoin instead of Bitcoin is that Litecoin allowed us to perform more experiments for the same money. While Litecoin is identical to Bitcoin in terms of most technical aspects, including priority policy to choose transactions for mining, we acknowledge that Bitcoin's longer confirmation interval would impose longer confirmation times on all transactions, including max-rate transactions.

\noindent
{\bf Blocking Max-Rate Transactions}.
Cryptocurrency ecosystems could modify their mining software to block max-rate transactions. We note that funding transactions are indistinguishable from any other P2SH transaction and thus can not be profiled or blocked by non-conforming nodes. We also note that modifications to mining software that attempt to block the corresponding spending transactions would make the blockchain open to cluttering from unspent transactions, that would excessively pollute the UTXO. Further, the financial motivations to obtain the associated mining fees are also incompatible with such a modification, which has not been observed so far in the wild.

\noindent
{\bf Participation Incentives}.
None of the Satoshi blockchains provide incentives for full node participation. However, UWeb actually provides an incentive for its users to host full nodes, in order to protect their access privacy. Thus, by providing incentives for hosting full nodes, UWeb has the potential to strengthen Satoshi ecosystems.

\noindent
{\bf Can PIR Reduce Bandwidth Use?}
While private information retrieval (PIR) solutions could reduce the communication costs imposed on UWeb clients, they would also cue the censor that the nodes running them use UWeb. This is because the reduced communications of PIR solutions can be profiled by the censor, and will look quite distinguishable from the communications of a standard Satoshi node. Second, the censor will see that the blocks retrieved by a node contain UWeb content, whereas those not retrieved do not contain such content. For instance, the use of Bloom filters specific for UWeb transactions will be distinguishable in terms of their UWeb content from the Bloom filters used by regular financial clients.

\section{Conclusions}
\label{sec:conclusion}

We have shown that arbitrary data insertion in commercial Satoshi blockchains can be practical when leveraging max-rate transactions, the customized smart contract constructs that we introduced. We have designed UWeb, a persistent data storage solution that builds on max-rate transactions to provide private and anonymous access to content and censorship resistance. We have implemented and evaluated UWeb on experiments with writing 268.21 MB of data in the Litecoin mainnet, and have shown that it achieves a 2-3 order of magnitude improvement on the storage throughput and blockchain utilization of state-of-the-art solutions. While our experiments broke Litecoin's daily block size lifetime record, they had only short-term effects on its ecosystem.

\section{Acknowledgments}

This research was supported by NSF grants CNS-2013671 and CNS-2114911, and CRDF grant G-202105-67826. This publication is based on work supported by a grant from the U.S. Civilian Research \& Development Foundation (CRDF Global). Any opinions, findings and conclusions or recommendations expressed in this material are those of the author(s) and do not necessarily reflect the views of CRDF Global.

\bibliographystyle{unsrt}  
\bibliography{bogdan,censorship,fs,fairswap,tithonus,secservices}

\appendix

\section{UWeb Analysis and Proofs}
\label{appendix:proofs}

We prove Claim~\ref{claim:standard}, i.e., that max-rate transactions are standard.

\begin{proof}
Our max-rate transaction $\tau$=$(v,f,I_T, O_T, w, l_o)$ ($\S$~\ref{sec:uweb:maxrate}) is standard since it satisfies the requirements of the reference client code~\cite{isStandardCode}.
Specifically, first {\it transaction weight check}: we choose $c$ = 59, so that the total size of $\tau$ is less than 100KB, see $\S$~\ref{sec:uweb:maxrate}. Second, {\it transaction inputs check}: the maximum size of our data storing script (see $\S$ \ref{sec:uweb:script}) is 1,568 bytes, thus smaller than 1,650B. Third, {\it number of OP\_RET operations check}: satisfied since our output list consists of only one OP\_RET output, see $\S$ \ref{sec:uweb:script}. Fourth, {\it transaction outputs check}: we do not use bare multi-sig outputs and since we use one OP\_RET output the dust check is ignored. Fifth, we satisfy the following conditions for the \texttt{isStandard()} function~\cite{isStandardCode}: (1)	{\it Solver function}: Our construct is of type \texttt{TX\_NULL\_DATA} as it uses one \texttt{OP\_RET} output ($\S$~\ref{sec:uweb:script}), (2) {\it Multi-Sig txn check}: Trivially satisfied, since we do not use multi-sig outputs, (3) {\it Null-transaction check}: We use a 0 length \texttt{OP\_RET} output, which is below 83B. We will include the full proof in the technical report.
\end{proof}

\noindent
{\bf Max-rate transactions prevent input modification attacks}.
Transaction validation for spending outputs guarantees the \textit{integrity} of the data contained in our input script. The hash lock constructs (lines 7-9 of Algorithm~\ref{fig:fattrans}) protect the data pushed (lines 2-4). Only the hash of the redeeming script can unlock the output while only the data pushed in the stack allows for an error-less script execution. This prevents the input modification attack of our spending transactions as only the intended data pushes will satisfy the redeeming script that spends our funding transaction output. Since the rules for standard transactions require that the stack is left with only one TRUE value, spurious data pushes preceding our data are not a concern.

\noindent
{\bf Max-rate transactions prevent output modification (rebind) attacks}.
Max-rate transactions make the output modification attacks impossible, by using  an OP\_RET output and a minimum fee rate in order to effectively prevent the adversary from \textit{stealing} dust value or decreasing the effective transaction fee rate. The attacker can not replace our OP\_RET output with even a dust value output as this would increase the size of the transaction (thus violating the transaction weight check~\cite{isStandardCode}) and any output value larger than zero would further reduce the fee rate to smaller than the minimum accepted for relaying (thus violating the transaction output check~\cite{isStandardCode} and restriction \ref{restriction:minimum_fees}, see $\S$~\ref{sec:model:system}).

\section{Integrity Attacks}
\label{appendix:attacks}

{\bf Output script modification attack}.
When receiving (over the gossip network) a transaction that does not include the operations OP\_CHECKSIG or OP\_CHECKSIGVERIFY in its input scripts, the adversary replaces the transaction's output addresses with his own. The lack of a signature verification involving the output scripts, implies that the result is a valid transaction. The adversary rushes to broadcast the modified transaction, in an attempt to get it mined ahead of the original transaction. If the modified transaction is mined before the original, the attacker succeeds and effectively steals the corresponding output values. Further, if data is stored in transaction outputs, the adversary can also modify this data, thus also achieves a data integrity attack.

\noindent
{\bf Input script modification attack}.
An attacker that receives a transaction with redeeming scripts that do not perform stack validation operations, e.g. via the sequences: OP\_HASH160 $<$hash160$>$ OP\_EQUAL or OP\_HASH160 $<$hash160$>$ OP\_EQUALVERIFY, can replace any data preceding the redeeming script and still obtain a valid input script, effectively achieving a data corruption attack. The attacker rushes to broadcast the modified transaction over the gossip network, in an effort to have it mined before the original.

\section{Block Utilization in the Tor-Writing Experiment}
\label{appendix:tor}

\begin{figure}
\centering
\includegraphics[width=0.89\columnwidth]{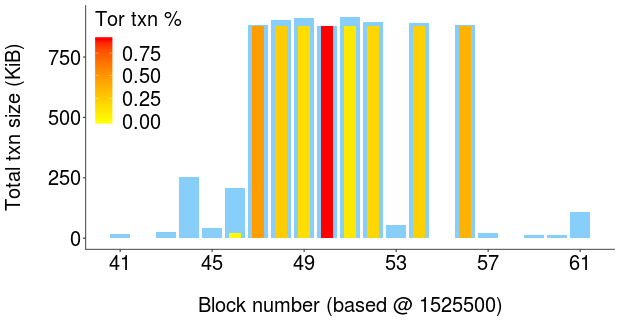}
\vspace{-10pt}
\caption{Impact on Litecoin blockchain of UWeb, Tor source code writing. Writing starts on block 46 and finishes on block 56. The block size increase is due to max-rate transactions, however, except for blocks 47 and 50, most Tor containing blocks contain many other small-size transactions.}
\label{fig:tor_writing_impact}
\vspace{-15pt}
\end{figure}

We zoom-in into block utilization results when using UWeb to write the Tor source code~\cite{torsource} 
compressed archive (6.4 MB).  Making Tor available to the world in the Litecoin blockchain required 17 funding and 73 spending transactions with a total cost of 0.0735 LTC (\$\ltcToUSD{0.0735}\result ~at the time of writing). Figure~\ref{fig:tor_writing_impact} shows the ability of UWeb and max-rate transactions to harness the blockchain capacity to the limit. The height of the bars represent the total transaction size per block. Our writing starts on block 1525546 and proceeds to block 1525556. We observe that blocks that do not contain our transactions (blocks smaller than 1525546 and bigger than 1525556) are mostly empty.

We note that block 1525550 consists of 10 max-rate transactions, each of 99,931B. Thus, our data is responsible for 90\% of the transactions in this block and takes up 99.96\% of the total block space. However, this usage of the Litecoin block did not result in a backup of financial transactions: for instance, block 1525553 that was mined in the middle of the experiment (Figure~\ref{fig:tor_writing_impact}), only contains financial transactions, and their total size is smaller than in blocks before our experiment, e.g., blocks 1525544 and 1525546. Further, our above experiments in the realnet (e.g., 140MB) and simulation (up to 1.14GB in Section~\ref{sec:evaluation:simulation}) show that writing at a much larger scale than the Tor writing experiment imposed no delays on financial transactions.

\section{UWeb-Written Censorship Resistant Systems \& Licenses}
\label{appendix:crslicenses}

\begin{table}[!ht]
\centering
\small
\caption{CRS source code permanently stored in the Litecoin Blockchain and their licenses that allow for free redistribution (Ordered by size).}
\vspace{-5pt}
\resizebox{0.49\textwidth}{!}{%
\textsf{
\begin{tabular}{l c c}
\toprule
Project Name & Redistribution License & Zip Size (KB)\\
\toprule
Stegotorus & 3-clause BSD &52000\\
Uproxy-client & Apache v2 &45000\\
Bit-smuggler & GPLv2 &42000\\
shadowsocks &  3-clause BSD &11000\\
lantern & Apache v2 &10000\\
tribler & GPLv3  &9700\\
Tor & Custom: Ok to redistribute & 6400 \\
telex & Custom: Ok to redistribute and Apache v2 &6000\\
twister-core & Custom: Ok to redistribute  &4500\\
infranet  & BSD  &2600\\
i2p.i2p & Public Domain  &2200\\
obfuscated-openssh & BSD or more free than BSD  &1900\\
ZeroNet & GPL v2 &1900\\
Dust & Custom: Ok to redistribute &1600\\
streisand & GPLv3 &1400\\
ipfs & MIT &978\\
firefly-proxy & Custom: 2-Clause BSD-like &930\\
Vuvuzela & GPL v3 &490\\
obfsproxy & Custom: Ok to redistribute &465\\
rubberhose & Custom: Ok to redistribute &392\\
WireGuard & GPL v2 &314\\
Whonix & GPLv3  &285\\
facebook-tunnel &  Apache v2 &264\\
cachebrowser &  MIT &230\\
go-packetflagon & Custom: 2-clause BSD-like  &179\\
codetalkertunnel & GPLv3 &169\\
gfw\_whitelist & MIT  &108\\
iodine & Custom: all permissions granted  &78\\
gfwlist & GPL v2.1  &73\\
govpn & GPLv3  &67\\
obfs4 & Custom: Ok to redistribute &67\\
GoodbyeDPI &  Apache v2  &42\\
marionette & Apache v2 &38\\
ChinaDNS & GPL v3 &37\\
Dat & Custom: 3-Clause BSD-like &37\\
gohop & GPLv3  &37\\
fteproxy & Apache v2 &28\\
antizapret & Creative Commons Legal Code &21\\
blockcheck & MIT &17\\
uproxy-server & Apache v2 &6.9\\
fwlite & GPL v3 &1.3\\
\bottomrule
\end{tabular}}}
\label{tables:tfs:licenses}
\vspace{-15pt}
\end{table}

\end{document}